\documentclass[runningheads]{llncs}

\usepackage{latexsym}
\usepackage{stmaryrd} %for bracket　　　　　　$\share{x}{1}{A}$
\usepackage{amsmath,amssymb,amsfonts}
\usepackage[noend]{algpseudocode}
\usepackage{algorithm}
\usepackage[dvipdfmx]{graphicx}
\usepackage{bm}
\usepackage{color}

\newcommand{\share}[3]{\llbracket {#1} \rrbracket_{#2}^{\mathsf{#3}}}% #1:element, #2:party, #3:scheme
\algnewcommand\algorithmicfunctionality{\textbf{Functionality:}}
\algnewcommand\Functionality{\item[\algorithmicfunctionality]}

\newcommand{\bit}{\{0, 1\}}
\newcommand{\View}[1]{\textsc{View}^\Pi_{{#1}}(\vec{x})}
\newcommand{\Output}{\textsc{Output}^\Pi(\vec{x})}
\newcommand{\algS}{\mathcal{S}}
\newcommand{\abs}[1]{\lvert #1 \rvert}
\newcommand{\Share}{\mathsf{Share}}
\newcommand{\Reconst}{\mathsf{Reconst}}
\newcommand{\Equality}{\mathsf{Equality}}
\newcommand{\TLU}{\mathsf{TLU}}
\newcommand{\MSNZB}{\mathsf{MSNZB}}
\newcommand{\Overflow}{\mathsf{Overflow}}
\newcommand{\Comparison}{\mathsf{Comparison}}
\newcommand{\BtoA}{\mathsf{B2A}}
\newcommand{\BXtoA}{\mathsf{BX2A}}
\newcommand{\BCtoA}{\mathsf{BC2A}}
\newcommand{\BCXtoA}{\mathsf{BCX2A}}
\newcommand{\Max}{\mathsf{Max}}
\newcommand{\Argmax}{\mathsf{Argmax}}
\newcommand{\Min}{\mathsf{Min}}
\newcommand{\Argmin}{\mathsf{Argmin}}

\newcommand{\bt}{\mathsf{bt}}
\newcommand{\ADD}{\textsf{ADD}}
\newcommand{\MULT}{\textsf{MULT}}
\newcommand{\XOR}{\textsf{XOR}}
\newcommand{\AND}{\textsf{AND}}
\newcommand{\NOT}{\textsf{NOT}}
\newcommand{\OR}{\textsf{OR}}
\newcommand{\Z}{\mathbb{Z}}

\begin{document}

\title{Communication-Efficient (Client-Aided) Secure Two-Party Protocols and Its Application}
\titlerunning{Communication-Efficient Secure Two-Party Protocols and Its Application}

\author{
	Satsuya Ohata\inst{1}
	%\orcidID{}
	\and
	Koji Nuida\inst{1,2}
	%\orcidID{}
}
\authorrunning{S. Ohata and K. Nuida}
\institute{
	National Institute of Advanced Industrial Science and Technology, Tokyo, Japan
	\email{satsuya.ohata@aist.go.jp}
	\and
	The University of Tokyo, Tokyo, Japan
	\email{nuida@mist.i.u-tokyo.ac.jp}
	}

\maketitle

\begin{abstract}
	Secure multi-party computation (MPC) allows a set of parties to compute a function jointly
	while keeping their inputs private.
	Compared with the MPC based on garbled circuits,
	some recent research results show that MPC based on secret sharing (SS) works at a very high speed.
	Moreover, SS-based MPC can be easily vectorized and achieve higher throughput.
    In SS-based MPC, however, we need many communication rounds for computing concrete protocols like 
    equality check, less-than comparison, etc.
	This property is not suited for large-latency environments like the Internet (or WAN).
	In this paper, we construct semi-honest secure communication-efficient two-party protocols.
	The core technique is {\textit {Beaver triple extension}}, which is a new tool for treating multi-fan-in gates,
	and we also show how to use it efficiently.
	We mainly focus on reducing the number of communication rounds, and our protocols also succeed in reducing
	the number of communication bits (in most cases).
	As an example, we propose a less-than comparison protocol (under practical parameters) with {\textit {three}} communication rounds.
	Moreover, the number of communication bits is also $38.4\%$ fewer.
	As a result, total online execution time is $56.1\%$ shorter than the previous work adopting the same settings.
	Although the computation costs of our protocols are more expensive than those of previous work,
	we confirm via experiments that such a disadvantage has small effects on the whole online performance
	in the typical WAN environments.
\end{abstract}

\section{Introduction}
\label{sec:introduction}

Secure multi-party computation (MPC) \cite{DBLP:conf/focs/Yao86,DBLP:conf/stoc/GoldreichMW87} 
allows a set of parties to compute a function $f$ jointly
while keeping their inputs private.
More precisely, the $N(\geq 2)$ parties, each holding private input $x_i$ for $i \in [1,N]$,
are able to compute the output $f(x_1,\cdots,x_N)$ without revealing their private inputs $x_i$.
Some recent research showed there are many progresses in the research on MPC based on secret sharing (SS) and
its performance is dramatically improved.
SS-based MPC can be easily vectorized and suitable for parallel executions.
We can obtain large throughput in SS-based MPC since we have no limit on the size of vectors.
This is a unique property on SS-based MPC, and it is compatible with the SIMD operations like
mini-batch training in privacy-preserving machine learning.
We cannot enjoy this advantage in the MPC based on garbled circuits (GC) or homomorphic encryption (HE).
The most efficient MPC scheme so far is three-party computation (3PC) based on $2$-out-of-$3$ SS
(e.g., \cite{DBLP:conf/ccs/ArakiFLNO16,DBLP:conf/crypto/ChidaGHIKLN18}).
In two-party computation (2PC), which is the focus of this paper, we need fewer hardware resources than 3PC.
Although it does not work at high speed since we need heavy pre-computation,
we can mitigate this problem by adopting slightly new MPC models
like client/server-aided models that we denote later.

In addition to the advantage as denoted above,
the amount of data transfer in online phase is also small in SS-based MPC than GC/HE-based one.
However, the number of communication rounds we need for computation is large in SS-based MPC.
We need one interaction between computing parties
when we compute an arithmetic multiplication gate or a boolean AND gate,
which is time-consuming when processing non-linear functions
since it is difficult to make the circuit depth shallow.
This is a critical disadvantage in real-world privacy-preserving applications since
there are non-linear functions we frequently use in practice
like equality check, less-than comparison, max value extraction, activation functions in machine learning, etc.
In most of the previous research, however, this problem has not been seriously treated.
This is because they assumed there is (high-speed) LAN connection between computing parties.
Under such environments, total online execution time we need for processing non-linear functions is small
even if we need many interactions between computing parties
since the communication latency is usually very short (typically $\leq0.5\mathrm{ms}$).
This assumption is somewhat strange in practice,
as the use of LAN suggests that MPC is executed on the network that is maintained by the same administrator/organization.
In that case, it is not clear if the requirement for SS that parties do not collude is held or not.
Hence, it looks more suitable to assume non-local networks like WAN.
However, large communication latency in WAN becomes the performance bottleneck in SS-based MPC.
We find by our experiments that the time caused by the communication latency occupies more than $99\%$ in some cases
for online total execution time.
To reduce the effect of the large communication latency,
it is important to develop SS-based MPC with fewer communication rounds.
In other words, we should put in work to make the circuit shallower
to improve the concrete efficiency of SS-based MPC.

\subsection{Related Work}
\label{subsec:relatedwork}

\paragraph{\bf{MPC Based on Secret Sharing}}
There are many research results on SS-based MPC.
For example, we have results on
highly-efficient MPC
(e.g., \cite{DBLP:conf/ccs/ArakiFLNO16,DBLP:conf/crypto/ChidaGHIKLN18}),
concrete tools or the toolkit
(e.g., \cite{DBLP:conf/tcc/DamgardFKNT06,DBLP:conf/pkc/NishideO07,DBLP:journals/ijisec/BogdanovNTW12,DBLP:conf/esorics/MoritaATONH18}),
mixed-protocol framework~\cite{DBLP:conf/ndss/Demmler0Z15,DBLP:conf/ccs/RiaziWTS0K18,DBLP:conf/ccs/MohasselR18},
application to privacy-preserving machine learning
or data analysis (e.g., \cite{DBLP:conf/sp/MohasselZ17,DBLP:conf/ccs/LiuJLA17,DBLP:conf/ccs/RiaziWTS0K18,DBLP:conf/ccs/MohasselR18,cryptoeprint:2019:695}),
proposal of another model for speeding up the pre-computation~\cite{DBLP:journals/popets/MohasselOR16,DBLP:conf/sp/MohasselZ17}, etc.
As denoted previously, however, 
we have not been able to obtain good experimental results for computing large circuits over WAN environments.
For example, \cite{DBLP:conf/sp/MohasselZ17} denoted the neural network training on WAN setting is not practical yet.
\paragraph{\bf{MPC Based on Garbled Circuit or Homomorphic Encryption}}
There are also many research results on GC/HE-based MPC.
For example, we have results on
the toolkit (e.g., \cite{DBLP:journals/tifs/LiuDCW16}),
encryption switching protocols~\cite{Kolesnikov2009HowTC,DBLP:conf/crypto/CouteauPP16},
application to private set intersection (e.g., \cite{DBLP:conf/eurocrypt/Pinkas0TY19})
or privacy-preserving machine learning (e.g., \cite{DBLP:conf/ndss/BostPTG15,DBLP:conf/icml/Gilad-BachrachD16,DBLP:journals/tifs/PhongAHWM18,DBLP:conf/crypto/BourseMMP18,DBLP:journals/popets/HesamifardTGW18,DBLP:conf/uss/JuvekarVC18}), etc.
Recently, we have many research results on
GC for more than three parties (e.g., \cite{DBLP:conf/ccs/MohasselRZ15,DBLP:conf/ccs/ZhuCS018}) and
Arithmetic GC (e.g., \cite{DBLP:journals/siamcomp/ApplebaumIK14}).
Note that it is difficult to improve the circuit size on standard boolean GC~\cite{DBLP:conf/eurocrypt/ZahurRE15},
which is a bottleneck on GC-based MPC.
Moreover, \cite{DBLP:conf/ccs/Ben-EfraimLO16,DBLP:conf/ccs/ByaliJPR18} proposed the GC-based MPC for WAN environments and
showed the benchmark using AES, etc.
Even if we adopt the most efficient GC~\cite{DBLP:conf/eurocrypt/ZahurRE15} with 128-bit security, however, we need to send at least 256-bit string per an AND gate.
This is two orders of magnitude larger than SS-based MPC. We construct the round-efficient protocol while keeping data traffic small.

\subsection{Our Contribution}
\label{subsec:ourcontribution}

There are two main contributions in this paper.
First, we propose the method for treating multi-fan-in gates in semi-honest secure SS-based 2PC and
show how to use them efficiently.
Second, we propose many round-efficient protocols and show their performance evaluations via experiments.
We explain the details of them as follows:
\begin{enumerate}
	\item
	We propose the method for treating multi-fan-in $\MULT/\AND$ gates over $\Z_{2^n}$ and
	some techniques for reducing the communication rounds of protocols.
	Our $N$-fan-in gates are based on the extension of Beaver triples,
	which is a technique for computing standard $2$-fan-in gates.
	In our technique, however, we have a disadvantage that the computation costs and the memory costs are exponentially increased by $N$;
	that is, we have to limit the size of $N$ in practice.
	On the other hand, we can improve the costs of communication.
	More concretely, we can compute arbitrary $N$-fan-in $\MULT/\AND$ with one communication round and
	the amount of data transfer is also improved.
	Moreover, we show performance evaluation results on above multi-fan-in gates via experiments.
	More concretely, see Sections~\ref{sec:coretool} and \ref{subsec:gateperformance}.
	\item
    We propose round-efficient protocols using multi-fan-in gates.
	We need fewer interactions for our protocols between computing parties in online phase than previous ones.
	When we use shares over $\Z_{2^{32}}$, compared with the previous work~\cite{DBLP:journals/ijisec/BogdanovNTW12},
	we reduce the communication rounds as follows:
	$\Equality:(5 \to 2)$, $\Comparison:(7 \to 3)$, and $\Max$ for $3$ elements:$(18 \to 4)$.
	Moreover, we show the performance evaluation results on our protocols via experiments.
	From our experiments, we find the computation costs for multi-fan-in gates and protocols based on them
	have small effects on the whole online performance in the typical WAN environments.
	We also implement an application (a privacy-preserving exact edit distance protocol for genome strings) using our protocols.
	More concretely, see Sections~\ref{sec:protocols}, \ref{subsec:protocolperformance}, and \ref{subsec:application}.
\end{enumerate}

\section{Preliminaries}
\label{sec:preliminaries}

\subsection{Syntax for Secret Sharing}
\label{subsec:syntaxss}
A $2$-out-of-$2$ secret sharing ($(2,2)$-SS) scheme over $\Z_{2^n}$ consists of two algorithms:
$\Share$ and $\Reconst$. $\Share$ takes as input $x \in \Z_{2^n}$,
and outputs $(\share{x}{0}{},\share{x}{1}{}) \in \Z_{2^n}^2$, 
where the bracket notation $\share{x}{i}{}$ denotes the share of the $i$-th party (for $i \in \{0,1\}$).
We denote $\share{x}{}{}=(\share{x}{0}{},\share{x}{1}{})$ as their shorthand. 
$\Reconst$ takes as input $\share{x}{}{}$, and outputs $x$.
For arithmetic sharing $\share{x}{}{A} = (\share{x}{0}{A}, \share{x}{1}{A})$ and
boolean sharing $\share{x}{}{B} = (\share{x}{0}{B}, \share{x}{1}{B})$,
we consider power-of-two integers $n$ (e.g. $n=64$) and $n=1$, respectively.

\subsection{Secure Two-Party Computation Based on (2,2)-Additive Secret Sharing}
\label{subsec:ss2pc}

Here, we explain how to compute arithmetic
$\ADD/\MULT$ gates on $(2,2)$-additive SS.
We use the standard $(2,2)$-additive SS scheme, defined by 
\begin{itemize}
	\item $\Share(x)$: randomly choose $r\in \Z_{2^n}$ and let $\share{x}{0}{A}=r$ and $\share{x}{1}{A}=x-r \in \Z_{2^n}$.	
	\item $\Reconst(\share{x}{0}{A}, \share{x}{1}{A})$: output $\share{x}{0}{A} + \share{x}{1}{A}$. 
\end{itemize}

We can compute fundamental operations; that is, $\ADD(x,y):=x+y$ and $\MULT(x,y):=xy$.
$\share{z}{}{} \gets \ADD(\share{x}{}{},\share{y}{}{})$ can be done locally
by just adding each party's shares on $x$ and on $y$. 
$\share{w}{}{} \gets \MULT(\share{x}{}{},\share{y}{}{})$ can be done in various ways.
We will use the standard method based on Beaver triples (BT)~\cite{DBLP:conf/crypto/Beaver91a}.
Such a triple consists of $\bt_0=(a_0,b_0,c_0)$ and $\bt_1=(a_1,b_1,c_1)$ such that 
$(a_0+a_1)(b_0+b_1)=(c_0+c_1)$.
Hereafter, $a$, $b$, and $c$ denote $a_0+a_1$, $b_0+b_1$, and $c_0+c_1$, respectively. 
We can compute these BT in offline phase.
In this protocol,
each $i$-th party $P_i$ ($i \in \bit$) can compute the multiplication share $\share{z}{i}{}=\share{xy}{i}{}$
as follows:
(1) $P_i$ first compute $(\share{x}{i}{}-a_i)$ and $(\share{y}{i}{}-b_i)$. 
(2) $P_i$ sends them to $P_{1-i}$.
(3) $P_i$ reconstruct $x'=x-a$ and $y'=y-b$.
(4) $P_0$ computes $\share{z}{0}{} = x'y' + x'b_0 + y'a_0 + c_0$ and $P_1$ computes $\share{z}{1}{} =x'b_1 + y'a_1 + c_1$.
Here, $\share{z}{0}{}$ and $\share{z}{1}{}$ calculated as above procedures are valid shares of $xy$;
that is, $\Reconst(\share{z}{0}{}, \share{z}{1}{}) = xy$.
We abuse notations and write the $\ADD$ and $\MULT$ protocols simply as
$\share{x}{}{}+\share{y}{}{}$ and
$\share{x}{}{} \cdot \share{y}{}{}$, respectively.
Note that similarly to the $\ADD$ protocol, 
we can also locally compute multiplication by constant $c$, denoted by $c \cdot \share{x}{}{}$. 

We can easily extend above protocols to boolean gates.
By converting $+$ and $-$ to $\oplus$ in arithmetic $\ADD$ and $\MULT$ protocols,
we can obtain $\XOR$ and $\AND$ protocols, respectively.
We can construct $\NOT$ and $\OR$ protocols from the properties of these gates.
When we compute $\NOT(\share{x}{0}{B}, \share{x}{1}{B})$, $P_0$ and $P_1$ output
$\lnot\share{x}{0}{B}$ and $\share{x}{1}{B}$, respectively.
When we compute $\OR(\share{x}{}{},\share{y}{}{})$, we compute $\lnot\AND(\lnot \share{x}{}{}, \lnot \share{y}{}{})$.
We abuse notations and write the $\XOR$, $\AND$, $\NOT$, and $\OR$ protocols simply as
$\share{x}{}{} \oplus \share{y}{}{}$,
$\share{x}{}{} \land \share{y}{}{}$,
$\lnot\share{x}{}{}$ (or $\overline{\share{x}{}{}}$), and
$\share{x}{}{} \lor \share{y}{}{}$, respectively.

\subsection{Semi-Honest Security and Client-Aided Model}
\label{subsec:semihonest}

In this paper, we consider simulation-based security notion in the presence of semi-honest adversaries
(for 2PC) as in~\cite{DBLP:books/cu/Goldreich2004}.
We show the definition in Appendix~\ref{appendix:semihonest}.
As described in~\cite{DBLP:books/cu/Goldreich2004}, 
composition theorem for the semi-honest model holds;
that is, any protocol is privately computed as long as its subroutines are
privately computed.

In this paper, we adopt client-aided model~\cite{DBLP:conf/sp/MohasselZ17,DBLP:conf/esorics/MoritaATONH18}
(or server-aided model~\cite{DBLP:journals/popets/MohasselOR16}) for 2PC.
In this model, a client (other than computing parties) generates and distributes shares of secrets.
Moreover, the client also generates and distributes some necessary BTs to the computing parties.
This improves the efficiency of offline computation dramatically
since otherwise computing parties would have to generate BTs by themselves jointly
via heavy cryptographic primitives like homomorphic encryption or oblivious transfer.
The only downside for this model is the restriction that
any computing party is assumed to not collude with the client who generates
BTs for keeping the security.

\section{Core Tools for Round-Efficient Protocols}
\label{sec:coretool}
In this section, we propose a core tool for round-efficient 2PC that we call
\lq\lq Beaver triple extension (BTE)''.
Moreover, we explain some techniques for pre-computation
to reduce the communication rounds in online phase.

\subsection{Example: 3-fan-in $\MULT/\AND$ via 3-Beaver Triple Extension}
\label{subsec:3gate}

Here, we explain the case of $3$-fan-in gates as an example.
We consider how to extend the mechanism of a $2$-fan-in $\MULT$ gate to a $3$-fan-in $\MULT$ gate ($3\texttt{-}\MULT$);
that is, we consider how to construct a special BT that cancels the terms
coming out from $(x-a)(y-b)(z-c)$ other than $xyz$.
We can obtain such one by extending the standard BT.
It consists of
$(a_0,b_0,c_0,d_0,e_0,f_0,g_0)$ for $P_0$ and $(a_1,b_1,c_1,d_1,e_1,f_1,g_1)$ for $P_1$
satisfying the conditions
$a_0+a_1=a, \cdots, g_0+g_1=g$, $ab=d$, $bc=e$, $ca=f$, and $abc=g$.
We call the above special BT as $3$-Beaver triple extension ($3$-BTE) in this paper.
We can compute the $3\texttt{-}\MULT$ using above $3$-BTE as follows:
\begin{enumerate}
	\item
	$P_i$ ($i \in \bit$) compute $(\share{x}{i}{}-a_i)$, $(\share{y}{i}{}-b_i)$, and $(\share{z}{i}{}-c_i)$.
	\item
	$P_i$ send them to another party.
	\item
	$P_i$ reconstruct $x'=x-a$, $y'=y-b$, and $z'=z-c$.
	\item
	$P_0$ computes
    $\share{w}{0}{} = x'y'z' + x'y'c_0 + y'z'a_0 + z'x'b_0
    + x'e_0 + y'f_0 + z'd_0 + g_0$ and
    $P_1$ computes
    $\share{w}{1}{} = x'y'c_1 + y'z'a_1 + z'x'b_1 + x'e_1 + y'f_1 + z'd_1 + g_1$.
\end{enumerate}
$\share{w}{0}{}$ and $\share{w}{1}{}$ are valid shares of $xyz$.
We can obviously construct a boolean $3$-BTE and $3$-fan-in $\AND$ gate ($3\texttt{-}\AND$)
by converting $+$ and $-$ to $\oplus$ in the $3\texttt{-}\MULT$ case and
also obtain $3$-fan-in $\OR$ gates ($3\texttt{-}\OR$).

\subsection{$N$-fan-in $\MULT/\AND$ via $N$-Beaver Triple Extension}
\label{subsec:Ngate}
% Here, we explain how to compute general $N$-fan-in gates.
\subsubsection{$N$-Beaver Triple Extension}
Let $N$ be a positive integer.
Let $\mathcal{M} = \Z_M$ for some $M$ (e.g., $M = 2^n$).
Write $[1,N] = \{1,2,\dots,N\}$.
%Let $\share{a}{0}{}$ and $\share{a}{1}{}$ denote two additive shares of a value
%$a \in \mathcal{M}$: $a = \share{a}{0}{} + \share{a}{1}{}$ in $\mathcal{M}$.
We define a client-aided protocol for generating $N$-BTE as follows:
\begin{enumerate}
	\item
	Client randomly chooses $\share{ a_{\{\ell\}} }{0}{}$ and $\share{ a_{\{\ell\}} }{1}{}$ from $\mathcal{M}$ ($\ell = 1,\dots,N$).
	Let $a_{\{\ell\}} \leftarrow \share{ a_{\{\ell\}} }{0}{} + \share{ a_{\{\ell\}} }{1}{}$.
	For each $I \subseteq [1,N]$ with $|I| \geq 2$, by setting $a_I \leftarrow \prod_{\ell \in I} a_{\{\ell\}}$,
	client randomly chooses $\share{ a_I }{0}{} \in \mathcal{M}$ and sets $\share{ a_I }{1}{} \leftarrow a_I - \share{ a_I }{0}{}$.
	\item
	Client sends all the $\share{ a_I }{0}{}$ to $P_0$ and all the $\share{ a_I }{1}{}$ to $P_1$.
\end{enumerate}
Note that, in the protocol above, the process of randomly choosing $\share{ a_I }{0}{}$
and then setting $\share{ a_I }{1}{} \leftarrow a_I - \share{ a_I }{0}{}$ is equivalent to randomly choosing
$\share{ a_I }{1}{}$ and then setting $\share{ a_I }{0}{} \leftarrow a_I - \share{ a_I }{1}{}$.
Therefore, the roles of $P_0$ and $P_1$ are symmetric.

\subsubsection{Multiplication Protocol}
For $\ell = 1,\dots,N$, let $(\share{ x_\ell }{0}{},\share{ x_\ell }{1}{})$
be given shares of $\ell$-th secret input value $x_\ell \in \mathcal{M}$.
The protocol for multiplication is constructed as follows:
\begin{enumerate}
	\item
	Client generates and distributes $N$-BTE $( \share{ a_I }{0}{} )_I$ and $( \share{ a_I }{1}{} )_I$
	to the two parties as described above.
	\item
	For $k = 0,1$, $P_k$ computes $\share{ x'_\ell }{k}{} \leftarrow \share{ x_\ell }{k}{} - \share{ a_{\{\ell\}} }{k}{}$ for $\ell = 1,\dots,N$ and sends those $\share{ x'_\ell }{k}{}$ to $P_{1-k}$.
	\item
	For $k = 0,1$, $P_k$ computes
	$x'_\ell \leftarrow \share{ x'_\ell }{1-k}{} + \share{ x'_\ell }{k}{}$ for $\ell = 1,\dots,N$.
	\item
	$P_0$ outputs $\share{ y }{0}{}$ given by
	\[
	\share{ y }{0}{}
	\leftarrow \prod_{\ell=1}^{N} x'_\ell + \sum_{\emptyset \neq I \subseteq [1,N]} \left( \prod_{\ell \in [1,N] \setminus I} x'_\ell \right) \share{ a_I }{0}{}
	\]
	while $P_1$ outputs $\share{ y }{1}{}$ given by
	\[
	\share{ y }{1}{}
	\leftarrow \sum_{\emptyset \neq I \subseteq [1,N]} \left( \prod_{\ell \in [1,N] \setminus I} x'_\ell \right) \share{ a_I }{1}{} \enspace.
	\]
\end{enumerate}

We can prove the correctness and semi-honest security of this protocol.
Due to the page limitation, we show the proofs in Appendix~\ref{appendix:BTEproof}.

\subsection{Discussion on Beaver Triple Extension}
\label{subsec:discussion_bte}
We can achieve the same functionality of $N\texttt{-}\MULT/\AND$ by using $2\texttt{-}\MULT/\AND$ multiple times
and there are some trade-offs between these two strategies.
% We show these trade-offs in Table \ref{table_n-fan-in} and
% discuss computation/communication costs in this section.
% In this table, memory cost means the number of elements we need for BT(E).
% A and M are the meaning of addition and multiplication, respectively.
% We assume that the circuit to compute $N\texttt{-}\AND$ using multiple $2\texttt{-}\AND$s is composed in a tree
% for reducing the circuit depth.
% As described later, our protocols based on BTE have intermediate property between
% SS-based protocols only using $2$-fan-in gates and MPCs based on garbled circuits.
% \begin{table*}[!t]
% 	\begin{center}
% 		\begin{small}
% 		\begin{tabular}{c|r|r|r|r}
% 											   & Memory Cost &        	                Comp. Cost & $\#$ of Comm. Bits &   $\#$ of Comm. Rounds \\ \hline
% 			%  Multiple Use of $2\texttt{-}\AND$ & 	     $6$ &	       					 $10A+10M$ & 			    $4$ &   				 $2$ \\
%              Multiple Use of $2\texttt{-}\AND$ &    $3(N-1)$ &					 $5(N-1)A+5(N-1)M$ & 		   $2(N-1)$ & $\lceil \log N \rceil$ \\ \hline
% 								    % $N$-$\AND$ & 		 $7$ & 						  	 $13A+20M$ & 			    $3$ & 				     $1$ \\
%                                     $N$-$\AND$ &   $2^{N}-1$ & $(2^{N+1}-3)A+(N \cdot 2^{N}-N-1)M$ & 			    $N$ & 				     $1$ \\
%         \end{tabular}
%         \vspace{2mm}
% 		\caption{The costs we need for memory, computation, and communication to achieve the functionality of $N\texttt{-}\AND$.}
% 		\label{table_n-fan-in}
% 	\end{small}
% 	\end{center}
% \end{table*}

\subsubsection{Memory, Computation, and Communication Costs}
In the computation of $N$-fan-in $\MULT/\AND$ using $N$-BTE, 
the memory consumption and computation cost increase exponentially with $N$.
Therefore, we have to put a restriction on the size of $N$ and
concrete settings change optimal $N$.
In this paper, we use $N\texttt{-}\MULT/\AND$ for $N \leq 9$
to construct round-efficient protocols.

$N$-fan-in $\MULT/\AND$ using $N$-BTE needs fewer communication costs.
Notably, the number of communication rounds of our protocol does not depend on $N$ and
this improvement has significant effects on practical performances in WAN settings.
Because of the problems on the memory/computation costs we denoted above, however, there is a limitation for the size of $N$.
When we use $L$-fan-in $\MULT/\AND$ ($L \leq N$) gates,
we need $\frac{\lceil \log N \rceil}{\lfloor \log L \rfloor}$ communication rounds
for computing $N$-fan-in $\MULT/\AND$.
When we set $L=8$, for example, we need two communication rounds to compute a $64$-fan-in $\AND$.

\subsubsection{Comparison with Previous Work}
Damg{\aa}rd et al.~\cite{DBLP:conf/tcc/DamgardFKNT06} also proposed how to compute $N$-fan-in gates in a round-efficient manner
using Lagrange interpolation.
Each of their scheme and ours has its merits and demerits.
Their scheme has an advantage over memory consumption and computational costs; that is,
their $N$-fan-in gates do not need exponentially large memory and computation costs.
On the other hand, their scheme needs two communication rounds to compute $N$-fan-in gates for any $N$ and
requires the share spaces to be $\Z_{p}$ ($p$: prime).
A 2PC scheme over $\Z_{2^n}$ is sometimes more efficient than one over $\Z_{p}$ when we implement them using
low-level language (e.g., C++) since we do not have to compute remainders modulo $2^n$ for all arithmetic operations.

\subsection{More Techniques for Reducing Communication Rounds}
\label{subsec:moretechnique}

\subsubsection{On Weights At Most One}
We consider the plain input $x$ that all bits are $0$, or only a single bit is $1$ and others are $0$.
For example, we consider $x=00100000$ and its boolean shares
$(\share{x}{0}{B},\share{x}{1}{B})=(10011011, 10111011)$.
We find these are correct boolean shares of $x$ since $\share{x}{0}{B} \oplus \share{x}{1}{B}=x$ holds.
In this setting, we can compute the share representing whether all the bits of $x$ are $0$ or not
without communications between $P_0$ and $P_1$.
More concretely, we can compute it by locally computing $\XOR$ for all bits on each share.
In the above example, $P_0$ and $P_1$ compute
$\bigoplus\share{x}{0}{B} = 1$ and $\bigoplus\share{x}{1}{B} = 0$, respectively.
$1 \oplus 0 = 1$ means there is $1$ in $x$.
This technique is implicitly used in the previous work~\cite{DBLP:journals/ijisec/BogdanovNTW12} for constructing
an arithmetic overflow detection protocol ($\Overflow$), which is an important building block for constructing
less-than comparison and more.
We show more skillful use of this technique for constructing $\Overflow$ to avoid heavy computation in our protocols.
More concretely, see Section \ref{subsec:protocol-overflow}.

\subsubsection{Arithmetic Blinding}
We consider the situation that two clients who have secrets also execute computation
(i.e., an input party is equal to a computing party), which is the different setting from client-aided 2PC.
In this case, $P_0$ and $P_1$ randomly split the secret $x$ and $y$ into $x_0,x_1$ and $y_0,y_1$, respectively.
Then $P_0$ sends $x_1$ to $P_1$ and $P_1$ sends $y_0$ to $P_0$.
If $P_0$ and $P_1$ previously obtain $a_0,b_0,c_0$ and $a_1,b_1,c_1$, respectively,
$P_0$ and $P_1$ can compute $\share{z}{}{}=xy$ via the standard multiplication protocol.
During this procedure, both $P_0$ and $P_1$ obtain $x-a$ and $y-b$.
Here, $P_0$ finds $a$ and $P_1$ finds $b$ since $P_0$ and $P_1$ know the value of $x$ and $y$, respectively.
Therefore, it does not matter if $P_0$ and $P_1$ previously know the corresponding values; that is,
$P_0$ can send $b_0$ to $P_1$ and $P_1$ can send $a_1$ to $P_0$ in the pre-computation phase.
This operation does not cause security problems.

By above pre-processing, $P_0$ and $P_1$ can directly send $x-a$ and $y-b$ in the multiplication protocol, respectively.
As a result, we can reduce the amount of data transfer in the multiplication protocol.
Note that in the setting that the input party is not equal to the computing party (e.g., standard client-aided 2PC),
this pre-processing does not work well since $P_0$ and $P_1$ do not have $x$ and $y$, respectively and
cannot compute $\share{z}{}{}=xy$ correctly.
Even in the client-aided 2PC setting, however, this situation appears in the
boolean-to-arithmetic conversion protocol.
More concretely, see Section \ref{subsec:protocol-b2a}.

\subsubsection{Trivial Sharing}
We consider the setting that an input party is not equal to a computing party,
which is the same one as standard client-aided 2PC.
In this situation, we can use the share $\share{b}{i}{}$ ($i \in \bit$) itself as a secret value for computations
by considering another party has the share $\share{0}{1-i}{}$.
Although we find this technique in the previous work~\cite{DBLP:journals/ijisec/BogdanovNTW12},
we can further reduce the communication rounds of two-party protocols by combining this technique and BTE.
More concretely, see Section \ref{subsec:protocol-b2a}.

\section{Communication-Efficient Protocols}
\label{sec:protocols}

In this section, we show round-efficient 2PC protocols using BTE and the techniques in Section \ref{subsec:moretechnique}.
For simplicity, in this section, we set a share space to $\Z_{2^{16}}$ and use $N$-fan-in gates ($N \leq 5$)
to explain our proposed protocols.
Although we omit the protocols over $\Z_{2^{32}}/\Z_{2^{64}}$ due to the page limitation,
we can obtain the protocols with the same communication rounds with $\Z_{2^{16}}$
by using $7$ or less fan-in $\AND$ over $\Z_{2^{32}}$ and $9$ or less fan-in $\AND$ over $\Z_{2^{64}}$.
We omit the correctness of the protocols adopting the same strategy in the previous work~\cite{DBLP:journals/ijisec/BogdanovNTW12}.

\subsection{Equality Check Protocol and Its Application}
\label{subsec:protocol-eqality}
An equality check protocol $\Equality(\share{x}{}{A}, \share{y}{}{A})$ outputs $\share{z}{}{B}$,
where $z=1$ iff $x=y$.
We start from the approach by \cite{DBLP:journals/ijisec/BogdanovNTW12} and focus on reducing communication rounds.
In $\Equality$, roughly speaking, we first compute $t=x-y$ and then check if all bits of $t$ are $0$ or not.
If all the bits of $t$ are $0$, it means $t=x-y=0$.
Although we can perform this functionality via $16\texttt{-}\OR$, we cannot directly execute such a large-fan-in $\OR$ gate.
We need $\log_{2}{16}=4$ communication rounds for the above procedure if we only use $2\texttt{-}\OR$ with a tree structure.
However, if we can use $4\texttt{-}\OR$, we can execute $\Equality$ with $\log_{4}{16}=2$ communication rounds.
We show our two-round $\Equality$ as in Algorithm~\ref{alg:equality}:
\begin{algorithm}[!t]
	\caption{Our Proposed $\Equality$}
	\label{alg:equality}
	\begin{algorithmic}[1]
		\Functionality $\share{z}{}{B} \gets \Equality(\share{x}{}{A}, \share{y}{}{A})$
		%\Require
		\Ensure $\share{z}{}{B}$, where $z=1$ iff $x=y$.
		\State $P_0$ and $P_1$ locally compute $\share{t}{0}{A} = \share{x}{0}{A} - \share{y}{0}{A}$ and $\share{t}{1}{A} = \share{y}{1}{A} - \share{x}{1}{A}$, respectively.
		\State $P_i$ ($i \in \bit$) locally extend $\share{t}{i}{A}$ to binary and see them as boolean shares; that is, $P_i$ obtain $[\share{t[15]}{i}{B}, \cdots, \share{t[0]}{i}{B}]$.
        \State $P_i$ compute $\share{t'[j]}{}{B} \gets 4\texttt{-}\OR(\share{t[4j]}{}{B}, \share{t[4j+1]}{}{B}, \share{t[4j+2]}{}{B}, \share{t[4j+3]}{}{B})$
        \Statex for $j \in [0,\cdots,3]$.
		\State $P_i$ compute $\share{t''}{}{B} \gets 4\texttt{-}\OR(\share{t'[0]}{}{B}, \share{t'[1]}{}{B}, \share{t'[2]}{}{B}, \share{t'[3]}{}{B})$.
		\State $P_i$ compute $\share{z}{}{B} = \lnot \share{t''}{}{B}$.
		\State \Return $\share{z}{}{B}$.
	\end{algorithmic}
\end{algorithm}
In this strategy, more generally, we need $\frac{\lceil \log n \rceil}{\lfloor \log L \rfloor}$
communication rounds for executing $\Equality$ when we set the share space to $\Z_{2^n}$ and use $N\texttt{-}\OR$ ($N \leq L$).

We can also obtain a round-efficient table lookup protocol $\TLU$ (or, $1$-out-of-$L$ oblivious transfer) using our $\Equality$.
We show the construction of three-round $\TLU$ in Appendix~\ref{appendix:subsec:tlu}.

\subsection{Overflow Detection Protocol and Applications}
\label{subsec:protocol-overflow}

An arithmetic overflow detection protocol $\Overflow$ has many applications and
is also a core building block of less-than comparison protocol.
The same as the approach by \cite{DBLP:journals/ijisec/BogdanovNTW12}, we construct $\Overflow$ via
the most significant non-zero bit extraction protocol $\MSNZB$.
We first explain how to construct $\MSNZB$ efficiently and then show two-round $\Overflow$.

A protocol for extracting the most significant non-zero bit ($\MSNZB(\share{x}{}{B}=[\share{x[15]}{}{B}, \cdots, \share{x[0]}{}{B}])$)
finds the position of the first \lq\lq$1$'' of the $x$ and outputs such a boolean share vector
$\share{z}{}{B}=[\share{z[15]}{}{B}, \cdots, \share{z[0]}{}{B}]$;
that is, for example, if $x=0010011100010000$, then $z=0010000000000000$.
To find the position of the first \lq\lq$1$'' in $x$ in a privacy-preserving manner,
we use a \lq\lq prefix-$\OR$'' operation~\cite{DBLP:journals/ijisec/BogdanovNTW12}.
In this procedure, we first replace further to the right bits than leftmost $1$ with $1$ via $2\texttt{-}\OR$ gates
and obtain $x'=0\cdots011\cdots1$.
Then, we compute $z = x' \oplus (x' \gg 1)$.
In this $\MSNZB$, we need four communication rounds since
$2\texttt{-}\OR$ runs four times even if we parallelize the processing.
Intuitively, we can construct two-round $\MSNZB$ via $4\texttt{-}\OR$; that is,
we compute multi-fan-in prefix-$\OR$ using $N\texttt{-}\OR$ ($N \leq 4$).
In this intuitive two-round $\MSNZB$, however, 
computation costs significantly increase since we have to compute $4\texttt{-}\OR$ many times.
Therefore, we consider how to reduce them while keeping the number of communication rounds.
We show our two-round $\MSNZB$ as in Algorithm \ref{alg:msnzb}.
% and Figure~\ref{fig:msnzb}:
\begin{algorithm}[!t]
	\caption{Our Proposed $\MSNZB$}
	\label{alg:msnzb}
	\begin{algorithmic}[1]
		\Functionality $\share{z}{}{B} \gets \MSNZB(\share{x}{}{B})$
		%\Require
		\Ensure $\share{z}{}{B}=[\share{z[15]}{}{B}, \cdots, \share{z[0]}{}{B}]$, where $z[j]=1$ for the largest value $j$ such that $x[j]=1$ and
		$z[k]=0$ for all $j \ne k$.
		\State $P_i$ ($i \in \bit$) set $\share{t[j]}{i}{B} = \share{x[j]}{i}{B}$ for $j \in [3,7,11,15]$. Then $P_i$ parallelly compute
		\Statex $\share{t[j]}{}{B} \gets 2\texttt{-}\OR(\share{x[j]}{}{B}, \share{x[j+1]}{}{B})$ for $j \in [2,6,10,14]$,
		\Statex $\share{t[j]}{}{B} \gets 3\texttt{-}\OR(\share{x[j]}{}{B}, \share{x[j+1]}{}{B}, \share{x[j+2]}{}{B})$ for $j \in [1,5,9,13]$, and
		\Statex $\share{t[j]}{}{B} \gets 4\texttt{-}\OR(\share{x[j]}{}{B}, \share{x[j+1]}{}{B}, \share{x[j+2]}{}{B}, \share{x[j+3]}{}{B})$ for $j \in [0,4,8,12]$.
		\State $P_i$ compute $\share{t'[j]}{i}{B} = \share{t[j]}{i}{B}$ for $j \in [3,7,11,15]$ and compute
		\Statex $\share{t'[j]}{i}{B} = \share{t[j]}{i}{B} \oplus \share{t[j+1]}{i}{B}$ for $j \in [0,1,2,4,5,6,8,9,10,12,13,14]$.
		\State $P_i$ locally compute $\share{s[j]}{i}{B} = \bigoplus_{k=4j}^{4j+3}\share{t'[k]}{i}{B}$ for $j \in [1,2,3]$.
		\State $P_i$ compute $\share{z[j]}{i}{B} = \share{t'[j]}{i}{B}$ for $j \in [12,\cdots,15]$. Then $P_i$ parallelly compute
		\Statex $\share{z[j]}{}{B} \gets 2\texttt{-}\AND(\share{t'[j]}{}{B}, \lnot \share{s[3]}{}{B})$ for $j \in [8,\cdots,11]$,
		\Statex $\share{z[j]}{}{B} \gets 3\texttt{-}\AND(\share{t'[j]}{}{B}, \lnot \share{s[2]}{}{B}, \lnot \share{s[3]}{}{B})$ for $j \in [4,\cdots,7]$, and
		\Statex $\share{z[j]}{}{B} \gets 4\texttt{-}\AND(\share{t'[j]}{}{B}, \lnot \share{s[1]}{}{B}, \lnot \share{s[2]}{}{B}, \lnot \share{s[3]}{}{B})$ for $j \in [0,\cdots,3]$.
		\State \Return $\share{z}{}{B} = [\share{z[15]}{}{B}, \cdots, \share{z[0]}{}{B}]$.
	\end{algorithmic}
\end{algorithm}
% \begin{figure}[tb]
% 	\centering
% 	\includegraphics[width=8.9cm]{fig_msnzb.pdf}
% 	\caption{An example of our $\MSNZB$ with $x=0000010100101101$.}
% 	\label{fig:msnzb}
% \end{figure}
In this construction, we first separate a bit string into some blocks and compute in-block $\MSNZB$.
Then, we compute correct $\MSNZB$ for $x$ via in-block $\MSNZB$.
In Algorithm~\ref{alg:msnzb}, we separate $16$-bit string uniformly into $4$ blocks
for avoiding the usage of large fan-in $\OR$.
This $\MSNZB$ is more efficient than the intuitive construction since we use fewer ($=4+4$) $4$-fan-in gates.

Based on the above $\MSNZB$, we can construct an arithmetic overflow detection protocol $\Overflow(\share{x}{}{A},k)$.
This protocol outputs $\share{z}{}{B}$,
where $z=1$ iff the condition $(\share{x}{0}{A} \mod 2^k + \share{x}{1}{A} \mod 2^k) \geq 2^{k}$ holds.
$\Overflow$ is an important building block of many other protocols that appear in the later of this section.
We also start from the approach by \cite{DBLP:journals/ijisec/BogdanovNTW12}.
In their $\Overflow$, we check whether or not there exists $1$ in $u=(-\share{x}{1}{} \mod 2^k)$ at the same position of
$\MSNZB$ on $d=((\share{x}{0}{} \mod 2^k) \oplus (-\share{x}{1}{} \mod 2^k))$.
Even if we apply our two-round $\MSNZB$ in this section, we need three communication rounds for their $\Overflow$
since we need one more round to check the above condition using $2\texttt{-}\AND$.
Here, we consider further improvements by combining $\MSNZB$ and $2\texttt{-}\AND$;
that is, we increase the fan-in of $\AND$ on the step 4 in Algorithm~\ref{alg:msnzb} and
push the computation of $2\texttt{-}\AND$ into that step as in Algorithm~\ref{alg:overflow}:
\begin{algorithm}[!t]
	\caption{Our Proposed $\Overflow$}
	\label{alg:overflow}
	\begin{algorithmic}[1]
		\Functionality $\share{z}{}{B} \gets \Overflow(\share{x}{}{A}, k)$
		\Ensure $\share{z}{}{B}$, where $z=1$ iff $(\share{x}{0}{A} \mod 2^k) + (\share{x}{1}{A} \mod 2^k) \geq 2^{k}$.
        \State $P_0$ locally extends ($\share{x}{0}{A} \mod 2^k$) to binary and obtains
        \Statex $\share{d}{0}{B}=[\share{d[15]}{0}{B}, \cdots, \share{d[0]}{0}{B}]$. $P_1$ also locally extends ($-\share{x}{1}{A} \mod 2^k$) to binary
        \Statex and obtains $\share{d}{1}{B}=[\share{d[15]}{1}{B}, \cdots, \share{d[0]}{1}{B}]$.
		\State $P_i$ ($i \in \bit$) set $\share{t[j]}{i}{B} = \share{d[j]}{i}{B}$ for $j \in [3,7,11,15]$. Then $P_i$ parallelly compute
		\Statex $\share{t[j]}{}{B} \gets 2\texttt{-}\OR(\share{d[j]}{}{B}, \share{d[j+1]}{}{B})$ for $j \in [2,6,10,14]$,
		\Statex $\share{t[j]}{}{B} \gets 3\texttt{-}\OR(\share{d[j]}{}{B}, \share{d[j+1]}{}{B}, \share{d[j+2]}{}{B})$ for $j \in [1,5,9,13]$, and
		\Statex $\share{t[j]}{}{B} \gets 4\texttt{-}\OR(\share{d[j]}{}{B}, \share{d[j+1]}{}{B}, \share{d[j+2]}{}{B}, \share{d[j+3]}{}{B})$ for $j \in [0,4,8,12]$.
        \State $P_i$ compute $\share{t'[j]}{i}{B} = \share{t[j]}{i}{B}$ for $j \in [3,7,11,15]$ and compute 
        \Statex $\share{t'[j]}{i}{B} = \share{t[j]}{i}{B} \oplus \share{t[j+1]}{i}{B}$ for $j \in [0,1,2,4,5,6,8,9,10,12,13,14]$.
		\State $P_i$ locally compute $\share{w[j]}{i}{B} = \bigoplus_{k=4j}^{4j+3}\share{t'[k]}{i}{B}$ for $j \in [1,2,3]$.
		\State $P_0$ sets $\share{u[j]}{0}{B} = 0$ for $j \in [0,\cdots,15]$ and
		\Statex $P_1$ sets $\share{u[j]}{1}{B} = \share{d[j]}{1}{B}$ for $j \in [0,\cdots,15]$.
		\State $P_i$ parallelly compute
		\Statex $\share{v[j]}{}{B} \gets 2\texttt{-}\AND(\share{t'[j]}{}{B}, \share{u[j]}{}{B})$ for $j \in [12,\cdots,15]$,
		\Statex $\share{v[j]}{}{B} \gets 3\texttt{-}\AND(\share{t'[j]}{}{B}, \share{u[j]}{}{B}, \lnot \share{w[3]}{}{B})$ for $j \in [8,\cdots,11]$,
		\Statex $\share{v[j]}{}{B} \gets 4\texttt{-}\AND(\share{t'[j]}{}{B}, \share{u[j]}{}{B}, \lnot \share{w[2]}{}{B}, \lnot \share{w[3]}{}{B})$ for $j \in [4,\cdots,7]$, and
		\Statex $\share{v[j]}{}{B} \gets 5\texttt{-}\AND(\share{t'[j]}{}{B}, \share{u[j]}{}{B}, \lnot \share{w[1]}{}{B}, \lnot \share{w[2]}{}{B}, \lnot \share{w[3]}{}{B})$ for $j \in [0,\cdots,3]$.
		\State $P_i$ locally compute $\share{z}{i}{B} = \bigoplus_{\ell=0}^{15}\share{v[\ell]}{i}{B}$.
		\State $P_i$ compute $\share{z}{}{B} = \lnot \share{z}{}{B}$.
		\State If $\share{x}{1}{A}=0$, then $P_1$ locally computes $\share{z}{1}{B} = \share{z}{1}{B} \oplus 1$
		\State \Return $\share{z}{}{B}$.
	\end{algorithmic}
\end{algorithm}
In our $\Overflow$, we need a communication for the steps $2$ and $6$ in Algorithm~\ref{alg:overflow}
and succeed in constructing two-round $\Overflow$ using $N\texttt{-}\AND$  $(N \leq 5)$ over $\Z_{2^{16}}$.
If we set the share space to $\Z_{2^{32}}/\Z_{2^{64}}$, we need to use $N\texttt{-}\AND$ for $N \leq 7 / N \leq 9$ for constructing two-round $\Overflow$, respectively.
Moreover, in Appendix~\ref{appendix:1roundoverflow}, we show more round-efficient $\Overflow$.
Although we need more computation and data transfer than $\Overflow$ in this section,
we can compute $\Overflow$ with one communication round (for small share spaces in practice).

We have many applications of $\Overflow$. We show the concrete construction of
less-than comparison ($\Comparison$) in Appendix~\ref{appendix:subsec:comparison}, which is a building block of the maximum value extraction protocol.
In particular, thanks to the round-efficient $\Overflow$, we can obtain a three-round $\Comparison$.
Morita et al.~\cite{DBLP:conf/esorics/MoritaATONH18} proposed a constant (= five)-round $\Comparison$
using multi-fan-in gates that works under the shares over $\Z_p$~\cite{DBLP:conf/tcc/DamgardFKNT06}.
Our $\Comparison$ is more round-efficient than theirs under the parameters we consider in this paper.

\subsection{Boolean-to-Arithmetic Conversion Protocol and Extensions}
\label{subsec:protocol-b2a}

A boolean-to-arithmetic conversion protocol $\BtoA(\share{x}{}{B})$ outputs $\share{z}{}{A}$, where $z=x$.
In (1-bit) boolean shares, there are four cases; that is,
$(\share{x}{0}{B},\share{x}{1}{B}) = (0,0), \allowbreak (0,1),(1,0),(1,1)$.
Even if we consider these boolean shares as arithmetic ones, it works well in the first three cases;
that is, $0\oplus0 = 0+0$, $0\oplus1 = 0+1$, and $1\oplus0 = 1+0$.
However, $1 \oplus 1 \ne 1 + 1$ and we have to correct the output of this case.
Based on this idea and the technique in Section \ref{subsec:moretechnique} (trivial sharing),
\cite{DBLP:journals/ijisec/BogdanovNTW12} proposed the construction of $\BtoA$.
In their protocol, we use a standard arithmetic multiplication protocol and need one communication round.
In the setting of client-aided 2PC, however, $\BtoA$ satisfies the condition that
input party is equal to the computing party.
Therefore, we can apply the techniques in Section \ref{subsec:moretechnique} (arithmetic blinding) and
construct more efficient $\BtoA$ as in Algorithm \ref{alg:b2a}:
\begin{algorithm}[!t]
	\caption{Our Proposed $\BtoA$}
	\label{alg:b2a}
	\begin{algorithmic}[1]
		\Functionality $\share{z}{}{A} \gets \BtoA(\share{x}{}{B})$
		\Ensure $\share{z}{}{A}$, where $z=x$.
		\State In pre-computation phase, the client randomly chooses $a,b \in \Z_{2^{16}}$, computes $c=ab$,
		chooses a randomness $r \in \Z_{2^{16}}$, and sets $(c_0,c_1) = (r,c-r)$.
		Then the client sends $(a,c_0)$ and $(b,c_1)$ to $P_0$ and $P_1$, respectively.
		\State $P_i$ ($i \in \bit$) set $\share{x}{i}{A} = \share{x}{i}{B}$.
		\State $P_0$ computes $x' = \share{x}{0}{A} - a$ and $P_1$ computes
		\Statex $x'' = \share{x}{1}{A} - b$.
		Then they send them to each other.
		\State $P_0$ computes $\share{z}{0}{A} = \share{x}{0}{A} - 2(x'x'' + x'' \cdot a + c_0)$ and
		\Statex $P_1$ computes $\share{z}{1}{A} = \share{x}{1}{A} - 2(x' \cdot b + c_1)$.
		\State \Return $\share{z}{}{A}$
	\end{algorithmic}
\end{algorithm}
Although the number of communication rounds is the same as in \cite{DBLP:journals/ijisec/BogdanovNTW12},
our protocol is more efficient.
First, the data transfer in online phase is reduced from $2n$-bits to $n$-bits.
Moreover, the number of randomnesses we need in pre-computation is reduced from five to three, and
the data amount for sending from the client to $P_0$ and $P_1$ is reduced from $3n$-bits to $2n$-bits.

We can extend the above idea and obtain protocols like
$\BXtoA$: $\share{b}{}{B} \times \share{x}{}{A} = \share{bx}{}{A}$,
$\BCtoA$: $\share{b}{}{B} \times \share{c}{}{B} = \share{bc}{}{A}$, and
$\BCXtoA$: $\share{b}{}{B} \times \share{c}{}{B} \times \share{x}{}{A} = \share{bcx}{}{A}$.
These protocols are useful when we construct a round-efficient maximum value extraction protocol (and its variants) in Section~\ref{subsec:protocol-max}.

\subsubsection{$\BXtoA$: $\share{b}{}{B} \times \share{x}{}{A} = \share{bx}{}{A}$}
\label{subsubsec:bx2a}
We usually need to compute the multiplication of a boolean share $\share{b}{}{B}$ and
an arithmetic one $\share{x}{}{A}$
(e.g., $\TLU$ in Section~\ref{appendix:subsec:tlu}, ReLU function in neural networks).
We call this protocol $\BXtoA$ in this paper.
\cite{DBLP:conf/ccs/MohasselR18} proposed one-round $\BXtoA$ under the $(2,3)$-replicated SS,
such construction in 2PC has not been known.
By almost the same idea as $\BtoA$, we can construct one-round $\BXtoA$ in 2PC as follows:
\begin{enumerate}
	\item
	$P_i$ ($i \in \bit$) set $\share{b}{i}{A} = \share{b}{i}{B}$.
	\item
	$P_0$ sets $\share{b'}{0}{A} = \share{b}{0}{B}$ and $\share{b''}{0}{A} = 0$, and
	$P_1$ sets $\share{b'}{1}{A} = 0$ and $\share{b''}{1}{A} = \share{b}{1}{B}$.
	\item
	$P_i$ compute
	\[
	\begin{split}
		\share{s}{i}{A} &\gets 2\texttt{-}\MULT(\share{b}{}{A},\share{x}{}{A})\\
		\share{t}{i}{A} &\gets 3\texttt{-}\MULT(\share{b'}{}{A},\share{b''}{}{A},\share{x}{}{A}).
	\end{split}
	\]
	\item
	$P_i$ computes $\share{z}{i}{A} = \share{s}{i}{A} - 2 \share{t}{i}{A}$.
\end{enumerate}
Here, we denote this computation as $\share{bx - 2 b_0 b_1 x}{}{A}$.

\subsubsection{$\BCtoA$: $\share{b}{}{B} \times \share{c}{}{B} = \share{bc}{}{A}$}
\label{subsubsec:bc2a}
Almost the same idea as $\BXtoA$, we can compute $\share{b}{}{B} \times \share{c}{}{B} = \share{bc}{}{A}$
$(\BCtoA)$ with one communication round.
We use this protocol in $3\texttt{-}\Argmax/3\texttt{-}\Argmin$ in Section~\ref{subsec:protocol-max}.
We can construct one-round $\BCtoA$ by computing
\[
	\share{bc - 2 b_0 b_1 - 2 c_0 c_1 + 2 b_0 \overline{c_0} b_1 \overline{c_1} + 2 \overline{b_0} c_0 \overline{b_1} c_1}{}{A}.
\]
We need $2\texttt{-}\MULT$ and $4\texttt{-}\MULT$ for this protocol.

\subsubsection{$\BCXtoA$: $\share{b}{}{B} \times \share{c}{}{B} \times \share{x}{}{A} = \share{bcx}{}{A}$}
\label{subsubsec:bcx2a}
Almost the same idea as the above protocols, we can also compute
$\share{b}{}{B} \times \share{c}{}{B}  \times \share{x}{}{A} = \share{bc}{}{A}$
$(\BCXtoA)$ with one communication round.
We use this protocol in $\Max/\Min$ in Section~\ref{subsec:protocol-max}.
We can construct one-round $\BCtoA$ by computing
\[
	\share{bcx - 2 b_0 b_1 x - 2 c_0 c_1 x + 2 b_0 \overline{c_0} b_1 \overline{c_1} x + 2 \overline{b_0} c_0 \overline{b_1} c_1 x}{}{A}.
\]
We need $3\texttt{-}\MULT$ and $5\texttt{-}\MULT$ for this protocol.

\subsection{The Maximum Value Extraction Protocol and Extensions}
\label{subsec:protocol-max}
The maximum value extraction protocol $\Max(\share{\bm{x}}{}{A})$
outputs $\share{z}{}{A}$, where $z$ is the largest value in $\bm{x}$.
We first explain the case of $\Max$ for three elements ($3\texttt{-}\Max$),
which is used for computing edit distance, etc.
We denote a $j$-th element of $\bm{x}$ as $x[j]$; that is, $\bm{x}=[x[0],x[1],x[2]]$.

We start from a standard tournament-based construction.
If the condition $x[0] < x[1]$ holds, $x' = x[1]$. Otherwise, $x' = x[0]$.
By repeating the above procedure once more using $\share{x'}{}{A}$ and $\share{x[2]}{}{A}$,
we can extract the maximum value among $\bm{x}$.
In this strategy, we need $16~(=(6+1+1) \times 2)$ communication rounds, and
$8~(=(3+1) \times 2)$ communication rounds even if we apply our three-round $\Comparison$ (in Section~\ref{subsec:protocol-overflow})
and $\BXtoA$ (in Section~\ref{subsec:protocol-b2a}).
This is mainly because we cannot parallelly execute $\Comparison$.
To solve this disadvantage, we first check the magnitude relationship for all elements using $\Comparison$.
Then we extract the maximum value.
Based on these ideas, we show our $3\texttt{-}\Max$ as in Algorithm~\ref{alg:3-max}:
\begin{algorithm}[!t]
	\caption{Our Proposed $3\texttt{-}\Max$}
	\label{alg:3-max}
	\begin{algorithmic}[1]
		\Functionality $\share{z}{}{A} \gets \Max(\share{\bm{x}}{}{A})$
		\Ensure $\share{z}{}{A}$, where $z$ is the largest element in $\bm{x}$.
		\State $P_i$ ($i \in \bit$) parallelly compute
		\Statex $\share{c_{01}}{}{B} \gets \Comparison(\share{x[0]}{}{A}, \share{x[1]}{}{A})$,
		\Statex $\share{c_{02}}{}{B} \gets \Comparison(\share{x[0]}{}{A}, \share{x[2]}{}{A})$, and
		\Statex $\share{c_{12}}{}{B} \gets \Comparison(\share{x[1]}{}{A}, \share{x[2]}{}{A})$.
		\State $P_i$ compute $\share{c_{10}}{i}{B} = \lnot \share{c_{01}}{i}{B}$, $\share{c_{20}}{i}{B} = \lnot \share{c_{02}}{i}{B}$, and $\share{c_{21}}{i}{B} = \lnot \share{c_{12}}{i}{B}$.
		\State $P_i$ parallelly compute
		\Statex $\share{t[0]}{i}{A} \gets \BCXtoA(\share{c_{10}}{}{B}, \share{c_{20}}{}{B}, \share{x[0]}{}{A})$,
		\Statex $\share{t[1]}{i}{A} \gets \BCXtoA(\share{c_{01}}{}{B}, \share{c_{21}}{}{B}, \share{x[1]}{}{A})$, and
		\Statex $\share{t[2]}{i}{A} \gets \BCXtoA(\share{c_{02}}{}{B}, \share{c_{12}}{}{B}, \share{x[2]}{}{A})$.
		\State $P_i$ compute $\share{z}{i}{A} = \Sigma_{j=0}^{2}\share{t[j]}{i}{A}$.
		\State \Return $\share{z}{}{A}$.
	\end{algorithmic}
\end{algorithm}
Although the computation costs obviously increased,
this is four-round $3\texttt{-}\Max$ by applying our $\Comparison$ and $\BCXtoA$.

Based on the above idea, we can also obtain
the minimum value extraction protocol, 
argument of the maximum/minimum extraction protocols,
and (argument of) the maximum/minimum value extraction protocols with $N(>3)$ inputs.
We show the construction of these protocols in
Appendix~\ref{appendix:subsec:min}, Appendix~\ref{appendix:subsec:argmax}, and Appendix~\ref{appendix:subsec:nmax}, respectively.

\section{Performance Evaluation}
\label{sec:performance}

We demonstrate the practicality of our arithmetic/boolean gates and protocols.
We implemented 2PC simulators and performed all benchmarks on a single laptop computer with Intel Core i7-6700K 4.00GHz and 64GB RAM.
We implemented simulators using Python 3.7 with Numpy v1.16.2 and vectorized all gates/protocols.
% We employ standard $\texttt{os.urandom}$ function to perform cryptographic randomness generation in our experiments.
We assumed $10\mathrm{MB/s}$ $(=80000\mathrm{bits/ms})$ bandwidth and $40\mathrm{ms}$ RTT latency as typical WAN settings, and
calculate the data transfer time (DTT) and communication latency (CL) using these values.
We adopted the client-aided model; that is, we assumed in our experiments that
clients generate BTE in their local environment without using HE/OT.

\subsection{Performance of Basic Gates}
\label{subsec:gateperformance}

\begin{figure}[tb]
	\centering
	\includegraphics[width=12.2cm]{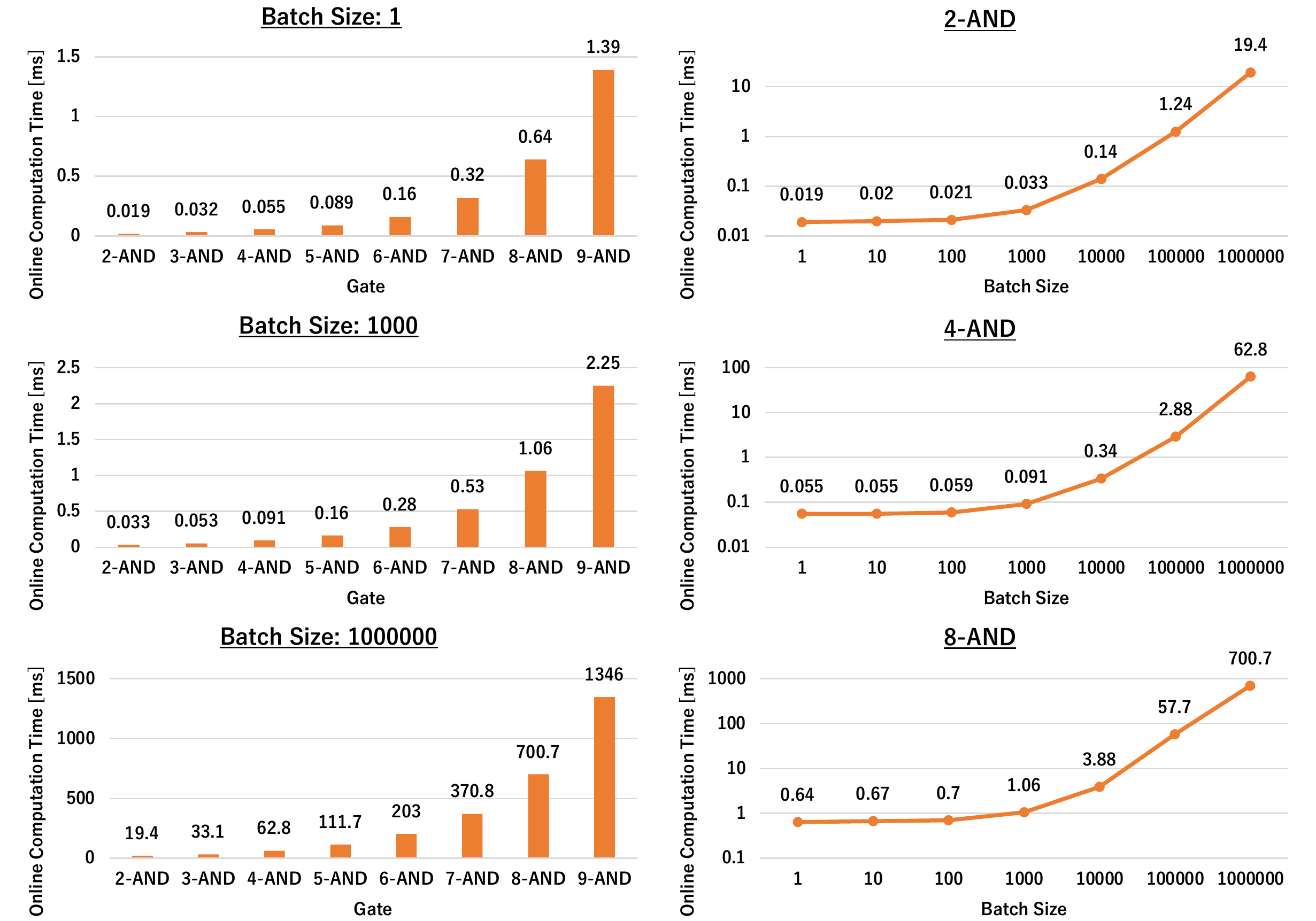}
	\caption{Relations between $N$ (fan-in number), batch size, and online computation time for $N\texttt{-}\AND$:
			 we show the relations between $N$ and online computation time with $1/1000/1000000$ batch (left),
			 and show the relations between batch size and online computation time for $2/4/8\texttt{-}\AND$ (right).}
	\label{fig:gate}
\end{figure}

Here we show experimental results on $N\texttt{-}\AND$.
We set $N=[2,\cdots,9]$ and $1$ to $10^6(=1000000)$ batch in our experiments.
Here we show the experimental results on the cases of $1/1000/1000000$ batch.
The experimental results on other cases ($10/100/10000/100000$ batch) are in Appendix~\ref{appendix:otherexpresults}.
The results are as in Table~\ref{table:gate} and Figure~\ref{fig:gate}:
\begin{table*}[tb]
	\begin{center}
		\begin{tiny}
			\begin{tabular}{c||r|r|r|r|r|r|r}
									 		&             \bf{pre-comp.} &          \bf{online comp.} & 		\bf{$\#$ of comm.} &          \bf{data trans.} & \bf{$\#$ of comm.} &                   \bf{comm.} & \bf{online total} \\
											&  \bf{time ($\mathrm{ms}$)} &  \bf{time ($\mathrm{ms}$)} & \bf{bits ($\mathrm{bit}$)} & \bf{time ($\mathrm{ms}$)} &        \bf{rounds} & \bf{latency ($\mathrm{ms}$)} & \bf{exec. time ($\mathrm{ms}$)} \\ \hline
											&           	     $0.015$ &              	  $0.019$ &      	 		       $2$ &     	$2.5\times10^{-5}$ &          	    $1$ &                    	  $40$ &                	      $40.0$ \\
				\bf{$2\texttt{-}\AND$}		&           	      $2.39$ &           	      $0.033$ & 	         $2\times10^3$ &  	  	$2.5\times10^{-2}$ &     	        $1$ &                    	  $40$ &           		          $40.1$ \\
											&           	      $2439$ &              	   $19.4$ &      	 	 $2\times10^6$ &      				$25.0$ &          	    $1$ &                    	  $40$ &                	      $84.4$ \\ \hline
											&           	     $0.041$ &              	  $0.032$ &      	       		   $3$ &       $3.75\times10^{-5}$ &          	    $1$ &                    	  $40$ &                	      $40.0$ \\
				\bf{$3\texttt{-}\AND$}	    &           	      $4.80$ &           	      $0.053$ & 	         $3\times10^3$ &  	   $3.75\times10^{-2}$ &     	        $1$ &                    	  $40$ &           		          $40.1$ \\
											&           	      $4899$ &              	   $33.1$ &      	     $3\times10^6$ &      				$37.5$ &          	    $1$ &                    	  $40$ &                	     $110.6$ \\ \hline
											&           	     $0.067$ &              	  $0.055$ &      	    		   $4$ &      	$5.0\times10^{-5}$ &          	    $1$ &                    	  $40$ &                	      $40.1$ \\
				\bf{$4\texttt{-}\AND$}	    &           	      $9.04$ &           	      $0.091$ & 	         $4\times10^3$ &  	  	$5.0\times10^{-2}$ &     	        $1$ &                    	  $40$ &           		          $40.1$ \\				
											&           	      $9383$ &              	   $62.8$ &      	     $4\times10^6$ &      				$50.0$ &          	    $1$ &                    	  $40$ &                	     $152.8$ \\ \hline
											&           	      $0.11$ &              	  $0.089$ &      	    		   $5$ &       $6.25\times10^{-5}$ &          	    $1$ &                    	  $40$ &                	      $40.1$ \\
				\bf{$5\texttt{-}\AND$}	    &           	      $17.2$ &           	       $0.16$ & 	         $5\times10^3$ &  	   $6.25\times10^{-2}$ &     	        $1$ &                    	  $40$ &           		          $40.2$ \\
											&           	     $17700$ &              	  $111.7$ &      	     $5\times10^6$ &      				$62.5$ &          	    $1$ &                    	  $40$ &                	     $214.2$ \\ \hline
											&           	      $0.20$ &              	   $0.16$ &      	     		   $6$ &     	$7.5\times10^{-5}$ &          	    $1$ &                    	  $40$ &                	      $40.2$ \\
				\bf{$6\texttt{-}\AND$}	    &           	      $33.0$ &           	       $0.28$ & 	         $6\times10^3$ &  	  	$7.5\times10^{-2}$ &     	        $1$ &                    	  $40$ &           		          $40.4$ \\
											&           	     $34059$ &              	  $203.0$ &      	     $6\times10^6$ &     				$75.0$ &          	    $1$ &                    	  $40$ &                	     $318.0$ \\ \hline
											&           	      $0.38$ &              	   $0.32$ &      	    		   $7$ &       $8.75\times10^{-5}$ &          	    $1$ &                    	  $40$ &                	      $40.3$ \\
				\bf{$7\texttt{-}\AND$}	    &           	      $64.3$ &           	       $0.53$ & 	         $7\times10^3$ &  	   $8.75\times10^{-2}$ &     	        $1$ &                    	  $40$ &           		          $40.6$ \\
											&           	     $66123$ &              	  $370.8$ &      	     $7\times10^6$ &      				$87.5$ &          	    $1$ &                    	  $40$ &                	     $498.3$ \\ \hline
											&           	      $0.76$ &              	   $0.64$ &      	    		   $8$ &      	$1.0\times10^{-4}$ &          	    $1$ &                    	  $40$ &                	      $40.6$ \\
				\bf{$8\texttt{-}\AND$}	    &           	     $125.1$ &           	       $1.06$ & 	         $8\times10^3$ &  	  	$1.0\times10^{-1}$ &     	        $1$ &                    	  $40$ &           		          $41.2$ \\
											&           	    $129553$ &              	  $700.7$ &      	     $8\times10^6$ &      			   $100.0$ &          	    $1$ &                    	  $40$ &                	     $840.7$ \\ \hline
											&           	      $1.63$ &              	   $1.39$ &      	    		   $9$ &      $1.125\times10^{-4}$ &          	    $1$ &                    	  $40$ &                	      $41.4$ \\
				\bf{$9\texttt{-}\AND$}	    &           	     $245.2$ &           	       $2.25$ & 	         $9\times10^3$ &  	  $1.125\times10^{-1}$ &     	        $1$ &                    	  $40$ &           		          $42.4$ \\
											&           	    $255847$ &              	   $1346$ &      	     $9\times10^6$ &      			   $112.5$ &          	    $1$ &                    	  $40$ &                	    $1498.5$ \\
			\end{tabular}
			\vspace{2mm}
			\caption{Evaluation on $N\texttt{-}\AND$ with $1$(upper)/$1000$(middle)/$1000000$(lower) batch.}
			\label{table:gate}
		\end{tiny}
	\end{center}
\end{table*}
We find
(1) the pre-computation time, online computation time, and data transfer time are exponentially growing up with respect to $N$;
(2) the dominant part in online total execution time is WAN latency especially in the case of small batch.
If we compute $N(>2)\texttt{-}\AND$ using multiple $2\texttt{-}\AND$ gates, we need two or more communication rounds.
Therefore, our scheme is especially suitable for the 2PC with relatively small batch (e.g., $\leq 10^5$)
as it yields low WAN latency.

\subsection{Performance of Our Protocols}
\label{subsec:protocolperformance}

\begin{figure*}[tb]
	\centering
	\includegraphics[width=12.2cm]{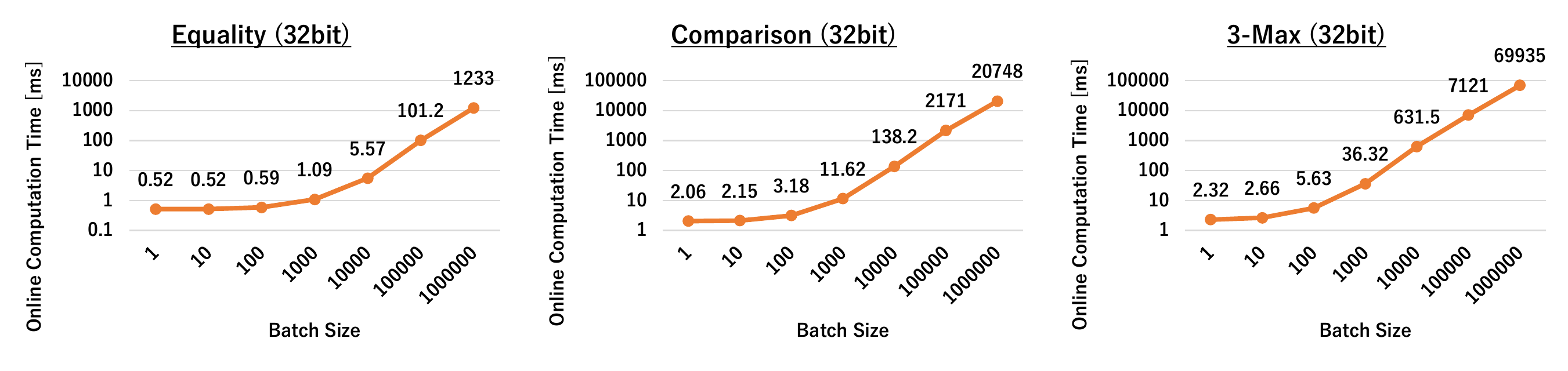}
	\caption{Relations between batch size and online computation/execution time of the protocols over $\Z_{2^{32}}$.}
	\label{fig:protocol1}
\end{figure*}

Here we show experimental results on our proposed protocols ($\Equality$, \allowbreak $\Comparison$, and $3\texttt{-}\Max$).
We implemented the baseline protocols~\cite{DBLP:journals/ijisec/BogdanovNTW12} and our proposed ones in Section \ref{sec:protocols}.
Same as the evaluation of $N\texttt{-}\AND$, we here show the results of our experiments over $\Z_{2^{32}}$ with $1/1000/1000000$ batch in
Table~\ref{table:protocol32} and Figure~\ref{fig:protocol1} (relations between batch size and online execution time).
Other results (protocols over $\Z_{2^{16}}$, $\Z_{2^{64}}$, and $\Z_{2^{32}}$ with other batch sizes) are in Appendix \ref{appendix:otherexpresults}.
\begin{table*}[tb]
	\begin{center}
		\begin{tiny}
			\begin{tabular}{c||r|r|r|r|r|r|r}
											&            \bf{pre-comp.} &         \bf{online comp.} & 		  \bf{$\#$ of comm.} &     	       \bf{data trans.} & \bf{$\#$ of comm.} &                   \bf{comm.} & \bf{online total} 			  \\
											& \bf{time ($\mathrm{ms}$)} & \bf{time ($\mathrm{ms}$)} & \bf{bits ($\mathrm{bit}$)} &    \bf{time ($\mathrm{ms}$)} &        \bf{rounds} & \bf{latency ($\mathrm{ms}$)} & \bf{exec. time ($\mathrm{ms}$)} \\ \hline
				\bf{$\Equality$}			&           	     $0.15$ &           	     $0.18$ & 	          		    $62$ & 	 	    $7.75\times10^{-4}$ &     	         $5$ &                        $200$ &           		      $200.2$ \\
				(1 batch)					&           $\mathbf{0.76}$ &           $\mathbf{0.52}$ &      	 	   $\mathbf{38}$ & $\mathbf{4.75\times10^{-4}}$ &       $\mathbf{2}$ &                $\mathbf{80}$ &                 $\mathbf{80.5}$ \\ \hline
				\bf{$\Comparison$}		    &           	      $1.5$ &           	     $0.54$ & 	          		   $970$ &  	    $1.21\times10^{-2}$ &     	         $7$ &                        $280$ &           		      $280.6$ \\
				(1 batch)					&            $\mathbf{3.9}$ &            $\mathbf{2.1}$ &      	      $\mathbf{712}$ &  $\mathbf{8.9\times10^{-3}}$ &       $\mathbf{3}$ &               $\mathbf{120}$ &                $\mathbf{122.1}$ \\ \hline
				\bf{$3\texttt{-}\Max$}	    &           	      $3.1$ &           	      $1.2$ & 	          		  $2196$ &  	    $2.75\times10^{-2}$ &     	        $18$ &                    	  $720$ &           		      $721.2$ \\
				(1 batch)					&            $\mathbf{9.7}$ &            $\mathbf{2.3}$ &      	     $\mathbf{3960}$ & $\mathbf{4.95\times10^{-2}}$ &       $\mathbf{4}$ &               $\mathbf{160}$ &                $\mathbf{162.3}$ \\ \hline\hline
				\bf{$\Equality$}			&           	     $74.7$ &           	     $0.61$ & 	          $62\times10^3$ & 	 	    			 $0.78$ &     	         $5$ &                        $200$ &           		      $201.4$ \\
				($10^3$ batch)				&          $\mathbf{500.5}$ &            $\mathbf{1.1}$ &    $\mathbf{38\times10^3}$ & 				$\mathbf{0.48}$ &       $\mathbf{2}$ &                $\mathbf{80}$ &                 $\mathbf{80.9}$ \\ \hline
				\bf{$\Comparison$}		    &           	     $1398$ &           	     $8.25$ & 	         $970\times10^3$ &  	                 $12.1$ &     	         $7$ &                        $280$ &           		      $300.4$ \\
				($10^3$ batch)				&           $\mathbf{2745}$ &           $\mathbf{11.6}$ &   $\mathbf{712\times10^3}$ &  			 $\mathbf{8.9}$ &       $\mathbf{3}$ &               $\mathbf{120}$ &                $\mathbf{140.5}$ \\ \hline
				\bf{$3\texttt{-}\Max$}	    &           	     $2891$ &           	     $17.5$ & 	        $2196\times10^3$ &  	    			 $27.5$ &     	        $18$ &                    	  $720$ &           		      $765.0$ \\
				($10^3$ batch)				&           $\mathbf{8635}$ &           $\mathbf{36.3}$ &  $\mathbf{3960\times10^3}$ & 				$\mathbf{49.5}$ &       $\mathbf{4}$ &               $\mathbf{160}$ &                $\mathbf{245.8}$ \\ \hline\hline
				\bf{$\Equality$}			&           	    $77574$ &           	    $761.4$ & 	          $62\times10^6$ & 	 	    		  	  $780$ &     	         $5$ &                        $200$ &           		       $1741$ \\
				($10^6$ batch)				&         $\mathbf{500617}$ &           $\mathbf{1233}$ &    $\mathbf{38\times10^6}$ & 			 	 $\mathbf{480}$ &       $\mathbf{2}$ &                $\mathbf{80}$ &                 $\mathbf{1793}$ \\ \hline
				\bf{$\Comparison$}		    &           	  $1445847$ &           	    $13895$ & 	         $970\times10^6$ &  	       	        $12100$ &     	         $7$ &                        $280$ &           		      $26275$ \\
				($10^6$0 batch)				&        $\mathbf{2799437}$ &          $\mathbf{20748}$ &   $\mathbf{712\times10^6}$ &  		 	$\mathbf{8900}$ &       $\mathbf{3}$ &               $\mathbf{120}$ &                $\mathbf{29768}$ \\ \hline
				\bf{$3\texttt{-}\Max$}	    &           	  $2956155$ &           	    $28252$ & 	        $2196\times10^6$ &  	    		 	$27500$ &     	        $18$ &                    	  $720$ &           		      $56472$ \\
				($10^6$ batch)				&        $\mathbf{8571664}$ &          $\mathbf{69935}$ &  $\mathbf{3960\times10^6}$ & 			   $\mathbf{49500}$ &       $\mathbf{4}$ &               $\mathbf{160}$ &               $\mathbf{119595}$ \\
									
			\end{tabular}
			\vspace{2mm}
			\caption{Evaluation of our protocols over $\Z_{2^{32}}$ for $1/10^3/10^6$ batches.
			In each cell, we show our experimental results on the baseline (upper) and ours (lower).}
			\label{table:protocol32}
		\end{tiny}
	\end{center}
\end{table*}
% \begin{figure*}[tb]
% 	\centering
% 	\includegraphics[width=12.3cm]{fig_protocol2.pdf}
% 	\caption{Throughput of baseline/our protocols over $\Z_{2^{32}}$:
% 			 $0$ means we cannot finish the protocol within that unit time (since sum of latency is larger than it).}
% 	\label{fig:protocol2}
% \end{figure*}
Same as the cases with $N\texttt{-}\AND$, WAN latency is the dominant part of the online total execution time.
In relatively small batch ($\leq 10^4$), 
all our protocols are faster than baseline ones in the online total execution time since ours require fewer communication rounds.
For example in $\Comparison$ with $1$ batch, we need more online computation time
than the baseline one ($0.54\textrm{ms} \to 2.1\textrm{ms}$).
However, communication costs of our $\Comparison$ are smaller than baseline one
(the number of communication rounds: $7 \to 3$, the number of communication bits: $970 \to 712$).
As a result, our $\Comparison$ is $56.1\%$ faster than baseline one
($280.6\textrm{ms} \to 122.1\textrm{ms}$) in our WAN settings.
As already mentioned, our protocols are not suitable for a (extremely) large batch
since the computation cost is larger than baseline ones.
% However, our experiments show that our protocol achieves higher throughput
% if the unit time is shorter than $1\mathrm{s}$.

\subsection{Application: Privacy-Preserving (Exact) Edit Distance}
\label{subsec:application}

We implemented a privacy-preserving edit distance protocol
using our protocols ($\Equality$, $\BtoA$, and $3\texttt{-}\Min$).
Unlike many previous works on approximate edit distance (e.g., \cite{DBLP:conf/ccs/0003T19}), here we consider the exact edit distance.
We computed an edit distance between two length-$L$ genome strings ($S_0$ and $S_1$) via standard dynamic programming (DP).
It appears four characters in the strings; that is, A, T, G, and C.
In DP-matrix, we fill the cell $x[i][j]$ by the following rule:
\[
	x[i][j] = 3\texttt{-}\Min([x[i-1][j]+1, x[i][j-1]+1, x[i-1][j-1]+e])
\]
Here, $e=0$ if the condition $S_0[i] = S_1[j]$ holds, and otherwise $e=1$.
We can compute $e$ using $\Equality$ (two rounds) and $\BtoA$ (one round).
To reduce the total online execution time, we calculate the edit distance as follows:
\begin{enumerate}
	\item
	To reduce the total communication rounds, we parallelly compute $e$ for all cells and store them in advance.
	Thanks to this procedure, we can avoid calculating $e$ every time when we fill cells.
	We only need three communication rounds for this step.
	\item
	Diagonal cells in DP-matrix are independent with each other.
	Therefore, we can parallelly compute these cells $x[d][0], x[d-1][1], \cdots, x[0][d]$ (for each $d$)
	to reduce the communication rounds.
\end{enumerate}
By applying the above techniques, we can compute exact edit distance for two length-$L$ strings
with $3+4(2L-1)=(8L-1)$ communication rounds.
We used the arithmetic shares and protocols over $\Z_{2^{16}}$ in our experiments.
The experimental results are as in Table~\ref{table:editdistance}:% and Figure~\ref{fig:editdistance}:
\begin{table*}[tb]
	\begin{center}
		\begin{tiny}
			\begin{tabular}{r||r|r|r|r|r}
				\bf{string} &           \bf{pre-comp.} &        \bf{online comp.} &     	\bf{data trans.} &                  \bf{comm.} & \bf{online total} 			    \\
				\bf{length}	& \bf{time ($\mathrm{s}$)} & \bf{time ($\mathrm{s}$)} & \bf{time ($\mathrm{s}$)} & \bf{latency ($\mathrm{s}$)} & \bf{exec. time ($\mathrm{s}$)} \\ \hline
						  4 &           	     $0.04$ &           	   $0.01$ & 	  $4.0\times10^{-4}$ &                      $1.24$ &           		      	 $1.25$ \\
						  8 &           	     $0.14$ &           	   $0.02$ & 	  $1.4\times10^{-3}$ &                      $2.52$ &           		      	 $2.54$ \\
						 16 &           	     $0.57$ &           	   $0.04$ & 	  $5.7\times10^{-3}$ &                      $5.08$ &           		      	 $5.13$ \\
						 32 &           	      $2.2$ &           	   $0.10$ & 	  $2.3\times10^{-2}$ &                      $10.2$ &           		      	 $10.3$ \\
						 64 &           	      $8.1$ &           	   $0.22$ & 	  $9.2\times10^{-2}$ &                      $20.4$ &           		      	 $20.7$ \\
						128 &           	     $33.4$ &           	   $0.54$ & 	  $3.7\times10^{-1}$ &                      $40.9$ &           		      	 $41.8$ \\
						256 &           	    $135.7$ &           	    $1.5$ & 	               $1.5$ &                      $84.9$ &           		      	 $84.9$ \\
						512 &           	    $534.1$ &           	    $4.8$ & 	               $5.9$ &                     $163.8$ &           		      	$174.5$ \\
					   1024 &           	     $2262$ &           	   $16.0$ & 	              $23.4$ &                     $327.6$ &           		      	$367.0$ \\
					\end{tabular}
			\vspace{2mm}
			\caption{Experimental results of privacy-preserving exact edit distance with $2^\ell$-length two strings ($\ell=[2,\cdots,10]$).}
			\label{table:editdistance}
		\end{tiny}
	\end{center}
\end{table*}
% \begin{figure}[tb]
% 	\centering
% 	\includegraphics[width=8.9cm]{fig_editdistance.pdf}
% 	\caption{Online execution time to compute exact edit distance in a privacy-preserving manner using our protocols.}
% 	\label{fig:editdistance}
% \end{figure}
As we can see from the experimental results, most of the online total execution time is occupied by the communication latency;
that is, GC-based approaches may be much faster than SS-based one in WAN environments.
However, if we would like to compute edit distances between many strings at the same time
(e.g., the situation that the client has one string and the server has 1000 strings, and the client would like to compute
edit distances between client's string and all of server's strings),
SS-based approach will be much faster than GC-based one.

\paragraph{\bf{Acknowledgements.}}
This work was partly supported by JST CREST JPMJCR19F6 and the Ministry of Internal Affairs and Communications Grant Number 182103105.

\bibliographystyle{splncs04}
\bibliography{ref_preprint_202001_WAN-MPC}

\newpage

\appendix

\section{Semi-Honest Security}
\label{appendix:semihonest}

Here, we recall the simulation-based security notion in the presence of semi-honest adversaries
(for 2PC) as in~\cite{DBLP:books/cu/Goldreich2004}.

\begin{definition}
\label{def:semihonest}
Let $f:(\bit^\ast)^2 \to(\bit^\ast)^2$ be a probabilistic $2$-ary functionality and 
$f_i(\vec{x})$ denotes the $i$-th element of $f(\vec{x})$ for 
$\vec{x}=(x_0, x_1)\in(\bit^\ast)^2$ and 
$i\in\{0,1\}$; 
$f(\vec{x}) = (f_0(\vec{x}), f_1(\vec{x}))$.
Let $\Pi$ be a $2$-party protocol to compute the functionality $f$.
The view of party $P_i$ for $i \in \{0,1\}$ during an execution
of $\Pi$ on input $\vec{x}=(x_0, x_1)\in(\bit^\ast)$ where 
$\abs{x_0} = \abs{x_1}$, denoted by $\View{i}$,
consists of $(x_i, r_i, m_{i,1}, \dots, m_{i, t})$, where 
$x_i$ represents $P_i$'s input, 
$r_i$ represents its internal random coins,
and $m_{i, j}$ represents 
the $j$-th message that $P_i$ has received.
The output of all parties after an execution of $\Pi$ on 
input $\vec{x}$ is denoted as $\Output$.
Then for each party $P_i$, 
we say that $\Pi$ {\em privately computes} $f$ in the presence of 
semi-honest corrupted party $P_i$ if there exists a probabilistic polynomial-time algorithm $\algS$ such that 
\begin{displaymath}
\{(\algS(i, x_i, f_i(\vec{x})), f(\vec{x}))\}
\equiv \{(\View{i}, \Output)\}
\end{displaymath}
where the symbol $\equiv$ means that the two probability distributions are statistically indistinguishable.
\end{definition}
As described in~\cite{DBLP:books/cu/Goldreich2004}, 
composition theorem for the semi-honest model holds;
that is, any protocol is privately computed as long as its subroutines are
privately computed.

\section{Correctness and Security of $N$-$\MULT/\AND$}
\label{appendix:BTEproof}

\subsection{Correctness of the Protocol}
We have
\[
\share{ y }{0}{} + \share{ y }{1}{}
= \prod_{\ell=1}^{N} x'_\ell + \sum_{\emptyset \neq I \subseteq [1,N]} \left( \prod_{\ell \in [1,N] \setminus I} x'_\ell \right) a_I \enspace.
\]
Since $a_I = \prod_{\ell \in I} a_{\{\ell\}}$, we have
\[
\begin{split}
\sum_{\emptyset \neq I \subseteq [1,N]} \left( \prod_{\ell \in [1,N] \setminus I} x'_\ell \right) a_I = ( x'_1 + a_{\{1\}} ) \cdots ( x'_N + a_{\{N\}} ) - x'_1 \cdots x'_N \enspace
\end{split}
\]
therefore (by noting that $x'_\ell = x_\ell - a_{\{\ell\}}$)
\[
\share{ y }{0}{} + \share{ y }{1}{}
= \prod_{\ell=1}^{N} ( x'_\ell + a_{\{\ell\}} )
= \prod_{\ell=1}^{N} x_\ell \enspace.
\]
Hence $\share{ y }{0}{}$ and $\share{ y }{1}{}$ form shares of $x_1 \cdots x_N$, as desired.

\subsection{Security Proof of the Protocol}
First we consider the security of the multiplication protocol against semi-honest $P_0$ (not colluding with Client).
Let $(\share{ x_\ell }{0}{},\share{ x_\ell }{1}{})$ ($\ell = 1,\dots,N$) be fixed input shares, and let $\zeta \in \mathcal{M}$.
We consider the conditional distribution of the view of $P_0$ for the case where the local output is $\share{ y }{0}{} = \zeta$.

The view of $P_0$ consists of $\share{ a_I }{0}{}$ for $\emptyset \neq I \subseteq [1,N]$ and $\share{ x'_\ell }{1}{}$
for $\ell = 1,\dots,N$ (note that the party uses no randomness in the protocol).
Let $\alpha_I$ for $\emptyset \neq I \subseteq [1,N]$ and $\gamma_\ell$ for $\ell = 1,\dots,N$ be elements of $\mathcal{M}$.
Let $E$ denote the corresponding event that $\share{ a_I }{0}{} = \alpha_I$ holds
for any $\emptyset \neq I \subseteq [1,N]$ and $\share{ x'_\ell }{1}{} = \gamma_\ell$ holds for any $\ell = 1,\dots,N$.
By the construction of the protocol, if the event $E$ occurs and moreover $\share{ y }{0}{} = \zeta$, then we have
\[
\zeta = \alpha_{[1,N]} + \varphi_0((\alpha_I)_{I \neq [1,N]},(\gamma_\ell)_\ell)
\]
where
\[
\varphi_0((\alpha_I)_{I \neq [1,N]},(\gamma_\ell)_\ell)
:= \prod_{\ell=1}^{N} \gamma_\ell + \sum_{\begin{subarray}{c} I \subseteq [1,N] \\ I \neq \emptyset,[1,N] \end{subarray}} \alpha_I \prod_{\ell \in [1,N] \setminus I} \gamma_\ell \enspace.
\]
This implies that the conditional probability $\Pr[ E \mid \share{ y }{0}{} = \zeta ]$ is $0$
if $\zeta \neq \alpha_{[1,N]} + \varphi_0((\alpha_I)_{I \neq [1,N]},(\gamma_\ell)_\ell)$.
We consider the other case where $\zeta = \alpha_{[1,N]} + \varphi_0((\alpha_I)_{I \neq [1,N]},(\gamma_\ell)_\ell)$.
Then the event $E$ implies $\share{ y }{0}{} = \zeta$.
Hence we have
\[
\Pr[ E \land \share{ y }{0}{} = \zeta ]
= \Pr[ E ] \enspace,
\]
therefore
\[
\Pr[ E \mid \share{ y }{0}{} = \zeta ]
= \Pr[ E ] / \Pr[ \, \share{ y }{0}{} = \zeta ] \enspace.
\]
Now the event $E$ occurs if and only if $\share{ a_I }{0}{} = \alpha_I$
for any $\emptyset \neq I \subseteq [1,N]$ and $\share{ a_{\{\ell\}} }{1}{} = \share{ x_\ell }{1}{} - \gamma_\ell$ for any $\ell = 1,\dots,N$.
As the choices of $\share{ a_I }{0}{}$'s and $\share{ a_{\{\ell\}} }{1}{}$'s are uniformly random and independent,
it follows that $\Pr[ E ]$ does not depend on $\alpha_I$'s and $\gamma_\ell$'s.
On the other hand, we have $\Pr[ \, \share{ y }{0}{} = \zeta ] = 1/|\mathcal{M}|$
(independent of $\alpha_I$'s and $\gamma_\ell$'s), as for any choice of $\share{ a_I }{0}{}$
for $I \neq \emptyset,[1,N]$ and of $\share{ x'_\ell }{1}{}$
there is precisely one possibility of $\share{ a_{[1,N]} }{0}{}$
that satisfies $\zeta = \share{ a_{[1,N]} }{0}{} + \varphi_0((\share{ a_I }{0}{})_{I \neq [1,N]},(\share{ x'_\ell }{1}{})_\ell)$.
Hence $\Pr[ E \mid \share{ y }{0}{} = \zeta ]$ is independent of $\alpha_I$'s and $\gamma_\ell$'s as well.

The argument above implies that, the distribution of the view of $P_0$ for fixed inputs and
given local output $\share{ y }{0}{} = \zeta$ is the uniform distribution on the set of tuples
$((\alpha_I)_I,(\gamma_\ell)_\ell)$ of elements of $\mathcal{M}$ satisfying
$\alpha_{[1,N]} + \varphi_0((\alpha_I)_{I \neq [1,N]},(\gamma_\ell)_\ell) = \zeta$.
The latter distribution can be sampled by freely choosing $\alpha_I$
for $I \neq [1,N]$ and $\gamma_\ell$ for $\ell = 1,\dots,N$ and then adjusting the value of $\alpha_{[1,N]}$.
Hence, the view of $P_0$ is efficiently and perfectly simulatable, implying the security against semi-honest $P_0$.
The argument showing the security against semi-honest $P_1$ is similar
(due to the aforementioned symmetry of the two parties in generating BTE),
where we use the function $\varphi_1$ instead of $\varphi_0$ given by
\[
\varphi_1((\alpha_I)_{I \neq [1,N]},(\gamma_\ell)_\ell)
:= \sum_{\begin{subarray}{c} I \subseteq [1,N] \\ I \neq \emptyset,[1,N] \end{subarray}} \alpha_I \prod_{\ell \in [1,N] \setminus I} \gamma_\ell \enspace.
\]
This concludes the security proof of the protocol.

\section{Applications and Extensions of Our Protocols}
\label{appendix:applicationsandextensions}

\subsection{Table Lookup}
\label{appendix:subsec:tlu}
We can also obtain a round-efficient table lookup protocol $\TLU$ (or, $1$-out-of-$L$ oblivious transfer) using our $\Equality$.
As shown in previous results, $\TLU$ is useful function in secure computation (e.g., \cite{DBLP:conf/ndss/DessoukyKS0ZZ17}).
Here, we consider the table of arithmetic keys/values with size $L$
(pairs of a $j$-th key $K_j$ and a $j$-th value $V_j$ for $j \in [0,\cdots,L-1]$).
We consider the situation that each computing party has shares of the table and a share of the index $\share{id}{i}{A}$ and
wants to obtain a share of the value $V_j$ where $id = K_j$.
To execute this protocol, we first check the equality of $id$ and $K_j$ for $j \in [0,\cdots,L-1]$ via $\Equality$.
Then, we extract $V_j$ using $\BXtoA$ (in Section~\ref{subsec:protocol-b2a}).
We only need three communication rounds for this $\TLU$.

\subsection{Less-Than Comparison}
\label{appendix:subsec:comparison}
To keep self-consistency of this paper, we explain how to construct a less-than comparison protocol $\Comparison(\share{x}{}{A}, \share{y}{}{A})$, which outputs
$\share{z}{}{B}$, where $z=1$ iff the condition $x<y$ holds.
The high-level construction of this protocol is completely the same as in~\cite{DBLP:journals/ijisec/BogdanovNTW12}; that is,
\begin{enumerate}
	\item
	$P_i$ $(i \in \bit)$ check whether the condition
	\[
		\share{x}{0}{A} \mod 2^{n-1} + \share{x}{1}{A} \mod 2^{n-1} > 2^{n-1}
	\]
	holds or not using $\Overflow$ and then compute
	$\share{x'}{}{B} = \share{\mathrm{of}_{x}}{}{B} \oplus \share{\mathrm{msb}_{x}}{}{B}$
	(and the same for $y$ and $d=x-y$, and obtain $\share{y'}{}{B}$ and $\share{d'}{}{B}$).
	Here, $\mathrm{of}_x$ denote the execution results of the above $\Overflow$ and
	$\mathrm{msb}_{x}$ denote the most significant bit of (binary expanded) $x$.
	We can extract the most significant bit of $x$, $y$, and $d$ via the above operations.
	\item
	$P_i$ compute
	\[
	\begin{split}
		\share{v}{}{B} &\gets 2\texttt{-}\AND((\share{x'}{}{B} \oplus \share{y'}{}{B}),\share{y'}{}{B})\\
		\share{w}{}{B} &\gets 2\texttt{-}\AND(\lnot(\share{x'}{}{B} \oplus \share{y'}{}{B}),\share{d'}{}{B}).
	\end{split}
	\]
	\item
	$P_i$ compute $\share{z}{}{B} = \share{v}{}{B} \oplus \share{w}{}{B}$.
\end{enumerate}

% \subsection{Arithmetic Rightshift}
% \label{appendix:subsec:rightshift}

% An arithmetic rightshift protocol $\Rightshift(\share{x}{}{A}, k)$ outputs $\share{z}{}{A}$, where $z = x \gg k$.
% We can construct three-round $\Rightshift$ using our $\Overflow$ as follows:
% \begin{enumerate}
% 	\item
% 	$P_i$ $(i \in \bit)$ compute
% 	\[
% 	\begin{split}
% 		\share{v}{}{B} \gets \Overflow(\share{x}{}{A}, k)\\
% 		\share{w}{}{B} \gets \Overflow(\share{x}{}{A}, n).
% 	\end{split}
% 	\]
% 	\item
% 	$P_i$ compute
% 	\[
% 	\begin{split}
% 		\share{v}{}{A} \gets \BtoA(\share{v}{}{B})\\
% 		\share{w}{}{A} \gets \BtoA(\share{w}{}{B}).
% 	\end{split}
% 	\]
% 	\item
% 	$P_i$ compute $\share{z}{i}{A} = \share{x}{i}{A} \gg k + \share{v}{i}{A} - 2^{n-k} \cdot \share{w}{i}{A}$.
% \end{enumerate}

% \subsection{Arithmetic Division}
% \label{appendix:subsec:division}
% An arithmetic division protocol (with private divisor) $\Division(\share{N}{}{A}, \share{D}{}{A})$ outputs $\share{z}{}{A}$,
% where $z = \lfloor \frac{N}{D} \rfloor$.
% We can improve the round complexity of $\Division$ since it calls $\Overflow$ and $\Rightshift$ many times
% as subroutines. Due to the lack of space, we omit the concrete construction of $\Division$ in this paper
% (see~\cite{DBLP:journals/ijisec/BogdanovNTW12}).

\subsection{The Minimum Value Extraction Protocol}
\label{appendix:subsec:min}

We can easily convert $\Max$ into $\Min$ by replacing the input order in step 1 in Algorithm~\ref{alg:3-max} and
obtain the minimum value extraction protocol for three elements $(3\texttt{-}\Min)$.
We use this $3\texttt{-}\Min$ for executing privacy-preserving exact edit distance protocol in Section~\ref{subsec:application}.

\subsection{$\Argmax$ and $\Argmin$}
\label{appendix:subsec:argmax}

We can easily obtain $\Argmax/\Argmin$ (by modifying $\Max/\Min$) as follows:
\begin{enumerate}
	\item
	We replace $\share{t[j]}{}{A} \gets \BCXtoA(\share{*}{}{B}, \share{**}{}{B}, \share{x}{}{A})$
	in Algorithm~\ref{alg:3-max} by $\share{t'[j]}{}{A} \gets \BCtoA(\share{*}{}{B}, \share{**}{}{B})$.
	\item
	$P_i$ compute $\share{z}{i}{A} = \Sigma_{j=0}^{2}(j \cdot \share{t'[j]}{i}{A})$ in the step 4 in Algorithm~\ref{alg:3-max}, .
\end{enumerate}
We can execute $\Argmax/\Argmin$ with three communication rounds.
We need fewer communication bits since we can avoid using $\BCXtoA$ in these protocols.
Note that in the above step 2, we need no interaction between computing parties since $j$ is public.

\subsection{$N\texttt{-}\Max/\Min$ for $N>3$}
\label{appendix:subsec:nmax}

Even in the cases of $\Max/\Min$ for four or more elements,
we can construct round-efficient $\Max/\Min$ with the same strategy as in Algorithm~\ref{alg:3-max}.
However, there are two points of notice as follows:
\begin{enumerate}
	\item
	In $N\texttt{-}\Max/\Min$, we need to (parallelly) execute
	$\Comparison$ $\frac{N(N-1)}{2}$ times.
	In the tournament-based strategy, we only need to execute $\Comparison$
	for $\lceil \log{}{N} \rceil$ times; that is, in our protocols,
	computation costs and the amount of communication bits rapidly increase with respect to $N$.
	\item
	For large $N$, we cannot directly use $\BCXtoA$ (or $\BCtoA$).
	Although we can construct the protocol like
	$\share{bcdx}{}{A} = \share{b}{}{B} \times \share{c}{}{B} \times \share{d}{}{B} \times \share{x}{}{A}$,
	we can easily imagine that the computation costs we need for such a protocol increase drastically.
	To avoid such a disadvantage, we should split the step 3 in Algorithm~\ref{alg:3-max} into some other protocols
	(e.g., $(N-1)\texttt{-}\AND$ and $\BXtoA$).
	This means we need more communication rounds to execute $N\texttt{-}\Max/\Min$ for large $N$.
\end{enumerate}

\section{One-Round $\Overflow$}
\label{appendix:1roundoverflow}

In this section, we explain another construction of $\Overflow$.
Although we need more computation and data transfer
than two-round $\Overflow$ in Section \ref{subsec:protocol-overflow},
we can compute the following $\Overflow$ with one communication round (for slightly small share spaces in practice).

\subsubsection{Protocol}
Let $\chi[P]$ denote a bit that is $1$ if the condition $P$ holds and $0$ otherwise.
Let $\Overflow(a,b;c) = \chi[a + b \geq c]$.
%Given a $2$-party arithmetic share $\AS{x} = (x_0,x_1)$ modulo $2^{\ell}$, the following protocol provides a Boolean share $\BS{\Overflow(x_0,x_1;2^{\ell})}$ of the bit $\Overflow(x_0,x_1;2^{\ell})$ detecting an overflow, where $\ell_1$ and $\ell_2$ are parameters with $\ell = \ell_1 + \ell_2$:
Here, $n_1$ and $n_2$ are parameters with $n=n_1+n_2$:
\begin{enumerate}
	\item
	\label{item:protocol_1}
	$P_i$ ($i \in \bit$) parses $\share{x}{i}{A} = y_i \mid\mid z_i$ where $y_i$ is the $n_1$ most significant bits of $x_i$ and 
	$z_i$ is the $n_2$ least significant bits of $x_i$.
	\item
	\label{item:protocol_2}
	For each $a_1 = 1,\dots,2^{n_1} - 1$,
	\begin{enumerate}
		\item
		$P_0$ sets $\alpha_0^{\langle a_1;1 \rangle} \gets \chi[ y_0 = a_1 ]$ and $\alpha_0^{\langle a_1;2 \rangle} \gets 0$.
		\item
		$P_1$ sets $\alpha_1^{\langle a_1;1 \rangle} \gets 0$ and $\alpha_1^{\langle a_1;2 \rangle} \gets \chi[ y_1 \geq 2^{n_1} - a_1 ]$.
	\end{enumerate}
	Let $\share{\alpha^{\langle a_1;j \rangle}}{}{B} = (\alpha_0^{\langle a_1;j \rangle}, \alpha_1^{\langle a_1;j \rangle})$ for $j = 1,2$.
	\item
	\label{item:protocol_3}
	For each $a_2 = 1,\dots,2^{n_2} - 1$ and $j = 0,\dots,n_1 - 1$, $P_0$ sets
	\[
	\beta_0^{\langle a_2;j \rangle} \gets
	\begin{cases}
	y_0[j] & \mbox{if } y_0 \neq 0 \mbox{ and } z_0 = a_2 \enspace,\\
	1 & \mbox{otherwise,}
	\end{cases}
	\]
	where $y_0[j]$ denotes the $j$-th bit of $y_0$.
	$P_1$ sets
	\[
	\beta_1^{\langle a_2;j \rangle} \gets
	\begin{cases}
	y_1[j] & \mbox{if } y_1 \neq 0 \mbox{ and } z_1 \geq 2^{n_2} - a_2 \enspace,\\
	1 & \mbox{otherwise.}
	\end{cases}
	\]
	Let $\share{\beta^{\langle a_2;j \rangle}}{}{B} = (\beta_0^{\langle a_2;j \rangle}, \beta_1^{\langle a_2;j \rangle})$.
	\item
	\label{item:protocol_4}
	For each $a_3 = 1,\dots,2^{n_2} - 1$,
	\begin{enumerate}
		\item
		$P_0$ sets $\gamma_0^{\langle a_3;1 \rangle} \gets \chi[ y_0 = 0 ]$, $\gamma_0^{\langle a_3;2 \rangle} \gets \chi[ y_0 = 2^{n_1} - 1 ]$, $\gamma_0^{\langle a_3;3 \rangle} \gets \chi[ z_0 = a_3 ]$, and $\gamma_0^{\langle a_3;4 \rangle} \gets 0$.
		\item
		$P_1$ sets $\gamma_1^{\langle a_3;1 \rangle} \gets \chi[ y_1 = 0 ]$, $\gamma_1^{\langle a_3;2 \rangle} \gets \chi[ y_1 = 2^{n_1} - 1 ]$, $\gamma_1^{\langle a_3;3 \rangle} \gets 0$, and $\gamma_1^{\langle a_3;4 \rangle} \gets \chi[ z_1 \geq 2^{n_2} - a_3 ]$.
	\end{enumerate}
	Let $\share{\gamma^{\langle a_3;j \rangle}}{}{B} = (\gamma_0^{\langle a_3;j \rangle}, \gamma_1^{\langle a_3;j \rangle})$ for $j = 1,2,3,4$.
	\item
	\label{item:protocol_5}
	Two parties execute the followings in parallel:
	For each $a_1 = 1,\dots,2^{n_1} - 1$, compute
	\[
	\share{b_1^{\langle a_1 \rangle}}{}{} \gets 2\texttt{-}\AND(\share{\alpha^{\langle a_1;1 \rangle}}{}{B}, \share{\alpha^{\langle a_1;2 \rangle}}{}{B})
	\]
	by using $2\texttt{-}\AND$.
	For each $a_2 = 1,\dots,2^{n_2} - 1$, compute
	\[
	\begin{split}
	\share{b_2^{\langle a_2 \rangle}}{}{B} \gets n_1\texttt{-}\AND(\share{\beta^{\langle a_2;0 \rangle}}{}{B}, \share{\beta^{\langle a_2;1 \rangle}}{}{B},~~~~~\\
	~~~~~~~~~~\dots, \share{\beta^{\langle a_2;n_1 - 1 \rangle}}{}{B} )
	\end{split}
	\]
	by using $n_1\texttt{-}\AND$.
	For each $a_3 = 1,\dots,2^{n_2} - 1$, compute
	\[
	\begin{split}
	\share{b_3^{\langle a_3 \rangle}}{}{B} \gets 4\texttt{-}\AND(\share{\gamma^{\langle a_3;1 \rangle}}{}{}, \share{\gamma^{\langle a_3;2 \rangle}}{}{B},~~~~~\\
	~~~~~~~~~~\share{\gamma^{\langle a_3;3 \rangle}}{}{B}, \share{\gamma^{\langle a_3;4 \rangle}}{}{B})
	\end{split}
	\]
	by using $4\texttt{-}\AND$.
	\item
	\label{item:protocol_6}
	$P_i$ locally compute
	\[
	\share{d}{i}{B} \gets \bigoplus_{a_1 = 1}^{2^{n_1} - 1} \share{b_1^{\langle a_1 \rangle}}{i}{B}
	\oplus \bigoplus_{a_2 = 1}^{2^{n_2} - 1} \share{b_2^{\langle a_2 \rangle}}{i}{B}
	\oplus \bigoplus_{a_3 = 1}^{2^{n_2} - 1} \share{b_3^{\langle a_3 \rangle}}{i}{B} \enspace.
	\]
	Then $P_i$ output the share $\share{d}{i}{B}$.
\end{enumerate}

All the steps except Step \ref{item:protocol_5} can be locally executed by each party.
Hence, in total, only $1$ round of communication is required which is spent during Step \ref{item:protocol_5}, where $(2^{n_1} - 1)$ $2\texttt{-}\AND$s, $(2^{n_2} - 1)$ $n_1\texttt{-}\AND$s, and $(2^{n_2} - 1)$ $4\texttt{-}\AND$s are performed in parallel.
For example, when $n = 15$ and $(n_1, n_2) = (8,7)$, these are $255$ $2\texttt{-}\AND$s, $127$ $8\texttt{-}\AND$s, and $127$ $4\texttt{-}\AND$s.

\subsubsection{Correctness}

First, we note that an overflow occurs modulo $2^{n}$ for $(x_0,x_1)$ if and only if, either an overflow occurs modulo $2^{n_1}$ for $(y_0,y_1)$, or $y_0 + y_1 = 2^{n_1} - 1$ and an overflow occurs modulo $2^{n_2}$ for $(z_0,z_1)$.
As the two events are disjoint, it follows that
\[
\begin{split}
\Overflow( x_0, x_1; 2^{n} )~~~~~~~~~~~~~~~~~~~~~~~~~~~~~~~~~~~~~~~~~~~~~~~~~~~~~~~~~~~~~~~~~~~~~~~~~~\\
~~~~~~~= \Overflow( y_0, y_1; 2^{n_1} ) \oplus ( \chi[ y_0 + y_1 = 2^{n_1} - 1 ] \land \Overflow( z_0, z_1; 2^{n_2} ) ) \enspace.
\end{split}
\]
Moreover, we have
\[
\chi[ y_0 + y_1 = 2^{n_1} - 1 ]
= \bigwedge_{j=0}^{n_1 - 1} ( y_0[j] \oplus y_1[j] ) \enspace,
\]
therefore $\Overflow( x_0, x_1; 2^{n} )$ is the XOR of $\Overflow( y_0, y_1; 2^{n_1} )$ and
\begin{equation}
	\label{eq:correctness_1}
	\left( \bigwedge_{j=0}^{n_1 - 1} ( y_0[j] \oplus y_1[j] ) \right)
	\land \Overflow( z_0, z_1; 2^{n_2} ) \enspace.
\end{equation}

For the term $\Overflow( y_0, y_1; 2^{n_1} )$, we note that the overflow occurs if and only if there is a (in fact, unique) $a_1 = 1,\dots,2^{n_1} - 1$ satisfying that $y_0 = a_1$ and $a_1 + y_1 \geq 2^{n_1}$ (i.e., $y_1 \geq 2^{n_1} - a_1$).
These $2^{n_1} - 1$ events are all disjoint.
In the protocol, the bit $\alpha^{\langle a_1;1 \rangle}$ is $1$ if and only if $y_0 = a_1$, and the bit $\alpha^{\langle a_1;2 \rangle}$ is $1$ if and only if $y_1 \geq 2^{n_1} - a_1$.
Therefore, we have
\[
\Overflow( y_0, y_1; 2^{n_1} )
= \bigoplus_{a_1 = 1}^{2^{n_1} - 1} \alpha^{\langle a_1;1 \rangle} \land \alpha^{\langle a_1;2 \rangle}
= \bigoplus_{a_1 = 1}^{2^{n_1} - 1} b_1^{\langle a_1 \rangle} \enspace.
\]
The same argument implies that, the bit in Eq.\eqref{eq:correctness_1} is equal to
\begin{tiny}
\begin{equation}
	\label{eq:correctness_2}
	\begin{split}
		&\left( \bigwedge_{j=0}^{n_1 - 1} ( y_0[j] \oplus y_1[j] ) \right)
		\land \bigoplus_{a_2 = 1}^{2^{n_2} - 1} \chi[ z_0 = a_2 ] \land \chi[ z_1 \geq 2^{n_2} - a_2 ] \\
		&= \bigoplus_{a_2 = 1}^{2^{n_2} - 1} \left( \left( \bigwedge_{j=0}^{n_1 - 1} ( y_0[j] \oplus y_1[j] ) \right) \land \chi[ z_0 = a_2 ] \land \chi[ z_1 \geq 2^{n_2} - a_2 ] \right) \enspace.
	\end{split}
\end{equation}
\end{tiny}

To decrease the depth of the circuit in Eq.\eqref{eq:correctness_2}, we consider to let $P_0$ modify the bits $y_0[j]$ in a way that the AND-term becomes $0$ if $z_0$ (known to $P_0$) is not equal to $a_2$.
Now observe that, unless $y_1 = 0$, at least one of the bits $y_1[j]$ is $1$, therefore the AND-term would become $0$ if all bits $y_0[j]$ were $1$.
Accordingly, instead of $y_0[j]$, we use a bit $y_0'[j]_{a_2}$ that is $y_0[j]$ if $z_0 = a_2$ and is $1$ if $z_0 \neq a_2$.
Then we have
\[
\begin{split}
&\left( \bigwedge_{j=0}^{n_1 - 1} ( y_0[j] \oplus y_1[j] ) \right) \land \chi[ z_0 = a_2 ] \land \chi[ z_1 \geq 2^{n_2} - a_2 ] \\
&= \left( \bigwedge_{j=0}^{n_1 - 1} ( y_0'[j]_{a_2} \oplus y_1[j] ) \right) \land \chi[ z_1 \geq 2^{n_2} - a_2 ] 
\end{split}
\]
unless $y_1 = 0$.
Similarly, we let $P_1$ modify the bits $y_1[j]$ in a way that the AND-term becomes $0$ if $z_1$ (known to $P_1$) is smaller than $2^{\ell_2} - a_2$.
Namely, instead of $y_1[j]$, we use a bit $y_1'[j]_{a_2}$ that is $y_1[j]$ if $z_1 \geq 2^{n_2} - a_2$ and is $1$ otherwise.
Then the same argument implies that
\[
\left( \bigwedge_{j=0}^{n_1 - 1} ( y_0'[j]_{a_2} \oplus y_1[j] ) \right) \land \chi[ z_1 \geq 2^{n_2} - a_2 ] = \bigwedge_{j=0}^{n_1 - 1} ( y_0'[j]_{a_2} \oplus y_1'[j]_{a_2} )
\]
unless $y_0 = 0$.
Summarizing, the bit in Eq.\eqref{eq:correctness_1} is equal to
\[
\bigoplus_{a_2 = 1}^{2^{n_2} - 1} \left( \bigwedge_{j=0}^{n_1 - 1} ( y_0'[j]_{a_2} \oplus y_1'[j]_{a_2} ) \right)
\]
unless $y_0 = 0$ or $y_1 = 0$.

Now we want to adjust the computation result in the case where $y_0 = 0$ or $y_1 = 0$.
Before doing that, we modify the computation further in order to simplify the situation: for $i = 0,1$, we change the bits $y_i'[j]_{a_2}$ in a way that it always becomes $1$ when $y_i = 0$.
The resulting bit is equal to $\beta_i^{\langle a_2;j \rangle}$ in the protocol, and the corresponding computation result
\begin{equation}
	\label{eq:correctness_3}
	\bigoplus_{a_2 = 1}^{2^{n_2} - 1} \left( \bigwedge_{j=0}^{n_1 - 1} ( \beta_0^{\langle a_2;j \rangle} \oplus \beta_1^{\langle a_2;j \rangle} ) \right)
	= \bigoplus_{a_2 = 1}^{2^{n_2} - 1} \left( \bigwedge_{j=0}^{n_1 - 1} \beta^{\langle a_2;j \rangle} \right)
	= \bigoplus_{a_2 = 1}^{2^{n_2} - 1} b_2^{\langle a_2 \rangle}
\end{equation}
is still equal to the bit in Eq.\eqref{eq:correctness_1} unless $y_0 = 0$ or $y_1 = 0$.
On the other hand, when $y_0 = 0$ or $y_1 = 0$, the bit in Eq.\eqref{eq:correctness_3} is equal to $0$, as now one of the two vectors $(\beta_i^{\langle a_2;0 \rangle}.\dots,\beta_i^{\langle a_2;n_1 - 1 \rangle})$ ($i = 0,1$) is $(1,1,\dots,1)$ while the other has at least one component being $1$.
When $y_0 = y_1 = 0$, the bit in Eq.\eqref{eq:correctness_1} is also equal to $0$ and hence is equal to the bit in Eq.\eqref{eq:correctness_3} as desired.
From now, we consider the other case where precisely one of $y_0$ and $y_1$ is equal to $0$; in the protocol, this is equivalent to $\gamma_0^{\langle a_3;1 \rangle} \oplus \gamma_1^{\langle a_3;1 \rangle} = 1$, i.e., $\gamma^{\langle a_3;1 \rangle} = 1$.
Under the condition, the bit in Eq.\eqref{eq:correctness_1} becomes $1$ if and only if the other $y_i$ which is not equal to $0$ is equal to $(11 \cdots 1)_2 = 2^{n_1} - 1$ (i.e., $\gamma^{\langle a_3;2 \rangle} = \gamma_0^{\langle a_3;2 \rangle} \oplus \gamma_1^{\langle a_3;2 \rangle} = 1$ in the protocol) and $\Overflow( z_0, z_1; 2^{n_2} ) = 1$.
By expanding the bit $\Overflow( z_0, z_1; 2^{n_2} )$ in the same way as the aforementioned case of $\Overflow( y_0, y_1; 2^{n_1} )$, it follows that the bit in Eq.\eqref{eq:correctness_1} is equal to
\[
\bigoplus_{a_3 = 1}^{2^{n_2} - 1} \gamma^{\langle a_3;1 \rangle} \land \gamma^{\langle a_3;2 \rangle} \land \gamma^{\langle a_3;3 \rangle} \land \gamma^{\langle a_3;4 \rangle}
= \bigoplus_{a_3 = 1}^{2^{n_2} - 1} b_3^{\langle a_3 \rangle}
\]
under the current condition.
Note that the bit above is $0$ when the current condition (i.e., precisely one of $y_0$ and $y_1$ is $0$) is not satisfied.

Summarizing the arguments, the bit in Eq.\eqref{eq:correctness_1} is equal to
\[
\bigoplus_{a_2 = 1}^{2^{n_2} - 1} b_2^{\langle a_2 \rangle}
\oplus \bigoplus_{a_3 = 1}^{2^{n_2} - 1} b_3^{\langle a_3 \rangle}
\]
in any case, therefore we have
\[
\begin{split}
\Overflow( x_0, x_1; 2^{n} )
&= \bigoplus_{a_1 = 1}^{2^{n_1} - 1} b_1^{\langle a_1 \rangle}
\oplus \bigoplus_{a_2 = 1}^{2^{n_2} - 1} b_2^{\langle a_2 \rangle}
\oplus \bigoplus_{a_3 = 1}^{2^{n_2} - 1} b_3^{\langle a_3 \rangle}\\
&= d
\end{split}
\]
as desired.
This completes the proof of correctness for the protocol.

\section{Other Experimental Results}
\label{appendix:otherexpresults}

Here we show the experimental results of our gates and protocols we omit in Section \ref{sec:performance}.

\begin{table*}[t]
	\begin{center}
		\begin{tiny}
			\begin{tabular}{c||r|r|r|r|r|r|r}
									 		&             \bf{pre-comp.} &          \bf{online comp.} & 		\bf{$\#$ of comm.} &          \bf{data trans.} & \bf{$\#$ of comm.} &                   \bf{comm.} & \bf{online total} \\
											&  \bf{time ($\mathrm{ms}$)} &  \bf{time ($\mathrm{ms}$)} & \bf{bits ($\mathrm{bit}$)} & \bf{time ($\mathrm{ms}$)} &        \bf{rounds} & \bf{latency ($\mathrm{ms}$)} & \bf{exec. time ($\mathrm{ms}$)} \\ \hline
											&           	     $0.037$ &              	  $0.020$ &      	 	 $2\times10^1$ &     	$2.5\times10^{-4}$ &          	    $1$ &                    	  $40$ &                	      $40.0$ \\
				\bf{$2\texttt{-}\AND$}		&           	      $0.24$ &           	      $0.021$ & 	         $2\times10^2$ &  	  	$2.5\times10^{-3}$ &     	        $1$ &                    	  $40$ &           		          $40.0$ \\
											&           	      $23.2$ &              	   $0.14$ &      	 	 $2\times10^4$ &     	$2.5\times10^{-1}$ &          	    $1$ &                    	  $40$ &                	      $40.4$ \\
											&           	     $243.1$ &              	    $1.2$ &      	 	 $2\times10^5$ &      				 $2.5$ &          	    $1$ &                    	  $40$ &                	      $43.7$ \\ \hline
											&           	     $0.085$ &              	  $0.033$ &      	     $3\times10^1$ &       $3.75\times10^{-4}$ &          	    $1$ &                    	  $40$ &                	      $40.0$ \\
				\bf{$3\texttt{-}\AND$}		&           	      $0.50$ &              	  $0.035$ &      	     $3\times10^2$ &       $3.75\times10^{-3}$ &          	    $1$ &                    	  $40$ &                	      $40.0$ \\
										    &           	      $46.3$ &           	       $0.21$ & 	         $3\times10^4$ &  	   $3.75\times10^{-1}$ &     	        $1$ &                    	  $40$ &           		          $40.6$ \\
											&           	     $489.8$ &              	    $1.9$ &      	     $3\times10^5$ &      				$3.75$ &          	    $1$ &                    	  $40$ &                	      $45.7$ \\ \hline
											&           	      $0.15$ &              	  $0.055$ &      	     $4\times10^1$ &      	$5.0\times10^{-4}$ &          	    $1$ &                    	  $40$ &                	      $40.1$ \\
				\bf{$4\texttt{-}\AND$}		&           	      $0.94$ &              	  $0.059$ &      	     $4\times10^2$ &      	$5.0\times10^{-3}$ &          	    $1$ &                    	  $40$ &                	      $40.1$ \\
										    &           	      $89.3$ &           	       $0.34$ & 	         $4\times10^4$ &  	  	$5.0\times10^{-1}$ &     	        $1$ &                    	  $40$ &           		          $40.8$ \\				
											&           	     $929.0$ &              	    $2.9$ &      	     $4\times10^5$ &      				 $5.0$ &          	    $1$ &                    	  $40$ &                	      $47.9$ \\ \hline
											&           	      $0.26$ &              	  $0.096$ &      	     $5\times10^1$ &       $6.25\times10^{-4}$ &          	    $1$ &                    	  $40$ &                	      $40.1$ \\
				\bf{$5\texttt{-}\AND$}	    &           	       $1.8$ &           	      $0.098$ & 	         $5\times10^2$ &  	   $6.25\times10^{-3}$ &     	        $1$ &                    	  $40$ &           		          $40.1$ \\
											&           	     $168.6$ &              	   $0.58$ &      	     $5\times10^4$ &       $6.25\times10^{-1}$ &          	    $1$ &                    	  $40$ &                	      $41.2$ \\				
											&           	      $1763$ &              	    $5.0$ &      	     $5\times10^5$ &      				$6.25$ &          	    $1$ &                    	  $40$ &                	      $51.3$ \\ \hline
											&           	      $0.49$ &              	   $0.17$ &      	     $6\times10^1$ &     	$7.5\times10^{-4}$ &          	    $1$ &                    	  $40$ &                	      $40.2$ \\
				\bf{$6\texttt{-}\AND$}	    &           	       $3.4$ &           	       $0.18$ & 	         $6\times10^2$ &  	  	$7.5\times10^{-3}$ &     	        $1$ &                    	  $40$ &           		          $40.2$ \\
											&           	     $327.8$ &              	    $1.0$ &      	     $6\times10^4$ &     	$7.5\times10^{-1}$ &          	    $1$ &                    	  $40$ &                	      $41.8$ \\
											&           	      $3379$ &              	   $13.1$ &      	     $6\times10^5$ &     				$7.50$ &          	    $1$ &                    	  $40$ &                	      $60.6$ \\ \hline
											&           	      $0.96$ &              	   $0.32$ &      	     $7\times10^1$ &       $8.75\times10^{-4}$ &          	    $1$ &                    	  $40$ &                	      $40.3$ \\
				\bf{$7\texttt{-}\AND$}	    &           	       $6.5$ &           	       $0.34$ & 	         $7\times10^2$ &  	   $8.75\times10^{-3}$ &     	        $1$ &                    	  $40$ &           		          $40.3$ \\
											&           	     $644.6$ &              	    $2.0$ &      	     $7\times10^4$ &       $8.75\times10^{-1}$ &          	    $1$ &                    	  $40$ &                	      $42.9$ \\
											&           	      $6564$ &              	   $27.4$ &      	     $7\times10^5$ &       				$8.75$ &          	    $1$ &                    	  $40$ &                	      $76.2$ \\ \hline
											&           	       $1.9$ &              	   $0.67$ &      	     $8\times10^1$ &      	$1.0\times10^{-3}$ &          	    $1$ &                    	  $40$ &                	      $40.7$ \\
				\bf{$8\texttt{-}\AND$}		&           	      $12.8$ &              	   $0.70$ &      	     $8\times10^2$ &      	$1.0\times10^{-2}$ &          	    $1$ &                    	  $40$ &                	      $40.7$ \\
										    &           	      $1274$ &           	        $3.9$ & 	         $8\times10^4$ &  	  				 $1.0$ &     	        $1$ &                    	  $40$ &           		          $44.9$ \\
											&           	     $12868$ &              	   $57.7$ &      	     $8\times10^5$ &      			    $10.0$ &          	    $1$ &                    	  $40$ &                	     $107.7$ \\ \hline
											&           	       $3.9$ &              	    $1.4$ &      	     $9\times10^1$ &      $1.125\times10^{-3}$ &          	    $1$ &                    	  $40$ &                	      $41.4$ \\
				\bf{$9\texttt{-}\AND$}	    &           	      $25.3$ &           	        $1.5$ & 	         $9\times10^2$ &  	  $1.125\times10^{-2}$ &     	        $1$ &                    	  $40$ &           		          $41.5$ \\
											&           	      $2538$ &              	    $9.9$ &      	     $9\times10^4$ &      			   $1.125$ &          	    $1$ &                    	  $40$ &                	      $51.0$ \\
											&           	     $25515$ &              	  $121.7$ &      	     $9\times10^5$ &      			   $11.25$ &          	    $1$ &                    	  $40$ &                	     $173.0$ \\
			\end{tabular}
			\vspace{2mm}
			\caption{Evaluation of $N\texttt{-}\AND$ with $10$(top)/$100$(second from the top)/$10000$(third from the top)/$100000$(bottom) batch.}
			\label{table:appendix_gate}
		\end{tiny}
	\end{center}
\end{table*}

\begin{table*}[!b]
	\begin{center}
		\begin{tiny}
			\begin{tabular}{c||r|r|r|r|r|r|r}
										&            \bf{pre-comp.} &         \bf{online comp.} & 		  \bf{$\#$ of comm.} &      	\bf{data trans.} & \bf{$\#$ of comm.} &                   \bf{comm.} & 				 \bf{online total} \\
										& \bf{time ($\mathrm{ms}$)} & \bf{time ($\mathrm{ms}$)} & \bf{bits ($\mathrm{bit}$)} & \bf{time ($\mathrm{ms}$)} &        \bf{rounds} & \bf{latency ($\mathrm{ms}$)} & \bf{exec. time ($\mathrm{ms}$)} \\ \hline
										&           	      $5.3$ &         	 	     $0.52$ & 	          $20\times10^1$ &  	 $4.75\times10^{-3}$ &     	          $2$ &                    	    $80$ &           		        $80.5$ \\
				\bf{$\Equality$}		&           	     $50.3$ &         	 	     $0.59$ & 	          $20\times10^2$ &  	 $4.75\times10^{-2}$ &     	          $2$ &                    	    $80$ &           		        $80.6$ \\
										&           	     $5003$ &         	 	      $5.6$ & 	          $20\times10^4$ &  				  $4.75$ &     	          $2$ &                    	    $80$ &           		        $90.4$ \\
										&          			$50051$ &          			$101.2$ &      	 	  $20\times10^5$ & 		  $4.75\times10^{1}$ &       		  $2$ &                			$80$ &                		   $228.7$ \\ \hline
										&           	     $28.1$ &           	      $2.2$ & 	         $712\times10^1$ &  	  $8.9\times10^{-2}$ &     	          $3$ &                    	   $120$ &           		       $122.3$ \\
				\bf{$\Comparison$}		&           	    $266.9$ &           	      $3.2$ & 	         $712\times10^2$ &  	  $8.9\times10^{-1}$ &     	          $3$ &                    	   $120$ &           		       $124.1$ \\
										&           	    $27810$ &           	    $138.2$ & 	         $712\times10^4$ &  	   $8.9\times10^{1}$ &     	          $3$ &                    	   $120$ &           		       $347.2$ \\
										&          		   $282130$ &          			 $2171$ &      	     $712\times10^5$ & 		   $8.9\times10^{2}$ &       		  $3$ &               		   $120$ &               		    $3181$ \\ \hline
										&           	     $83.5$ &           	      $2.7$ & 	        $3960\times10^1$ &  	 $2.75\times10^{-1}$ &     	          $4$ &                    	   $160$ &           		       $163.0$ \\
				\bf{$3\texttt{-}\Max$}	&           	    $841.3$ &           	      $5.6$ & 	        $3960\times10^2$ &  	    		  $2.75$ &     	          $4$ &                    	   $160$ &           		       $168.4$ \\
										&           	    $86345$ &           	    $631.5$ & 	        $3960\times10^4$ &  	  $2.75\times10^{2}$ &     	          $4$ &                    	   $160$ &           		        $1067$ \\
										&          		   $863023$ &      		    	 $7121$ &		    $3960\times10^5$ & 		  $2.75\times10^{3}$ &       		  $4$ &               		   $160$ &               		   $10031$ \\
			\end{tabular}
			\vspace{2mm}
			\caption{Evaluation of our protocols over $\Z_{2^{32}}$ with $10$(top)/$100$(second from the top)/$10000$(third from the top)/$100000$(bottom) batch.}
			\label{table:appendix_protocol32}
		\end{tiny}
	\end{center}
\end{table*}
\begin{table*}[t]
	\begin{center}
		\begin{tiny}
			\begin{tabular}{c||r|r|r|r|r|r|r}
										&            \bf{pre-comp.} &         \bf{online comp.} & 		  \bf{$\#$ of comm.} &      	\bf{data trans.} & \bf{$\#$ of comm.} &                   \bf{comm.} & 				 \bf{online total} \\
										& \bf{time ($\mathrm{ms}$)} & \bf{time ($\mathrm{ms}$)} & \bf{bits ($\mathrm{bit}$)} & \bf{time ($\mathrm{ms}$)} &        \bf{rounds} & \bf{latency ($\mathrm{ms}$)} & \bf{exec. time ($\mathrm{ms}$)} \\ \hline
										&           	     $0.17$ &         	 	     $0.17$ & 	          		    $20$ &  	  $2.5\times10^{-4}$ &     	          $2$ &                    	    $80$ &           		        $80.2$ \\
										&           	     $0.59$ &         	 	     $0.17$ & 	        $20\times10^{1}$ &  	  $2.5\times10^{-3}$ &     	          $2$ &                    	    $80$ &           		        $80.2$ \\
										&           	      $4.6$ &         	 	     $0.19$ & 	        $20\times10^{2}$ &  	  $2.5\times10^{-2}$ &     	          $2$ &                    	    $80$ &           		        $80.2$ \\
				\bf{$\Equality$}		&           	     $46.0$ &         	 	     $0.36$ & 	        $20\times10^{3}$ &  	  $2.5\times10^{-1}$ &     	          $2$ &                    	    $80$ &           		        $80.6$ \\
										&           	    $447.3$ &         	 	      $1.8$ & 	        $20\times10^{4}$ &  				   $2.5$ &     	          $2$ &                    	    $80$ &           		        $84.3$ \\
										&           	     $4591$ &         	 	     $30.5$ & 	        $20\times10^{5}$ &  	   $2.5\times10^{1}$ &     	          $2$ &                    	    $80$ &           		       $135.5$ \\
										&          			$45982$ &          			$376.9$ &      	 	$20\times10^{6}$ & 		   $2.5\times10^{2}$ &       		  $2$ &                			$80$ &                		   $706.9$ \\ \hline
										&           	     $0.95$ &           	     $0.98$ & 	          		   $280$ &  	  $3.5\times10^{-3}$ &     	          $3$ &                    	   $120$ &           		       $121.0$ \\
										&           	      $6.4$ &           	     $0.98$ & 	       $280\times10^{1}$ &  	  $3.5\times10^{-2}$ &     	          $3$ &                    	   $120$ &           		       $121.0$ \\
										&           	     $58.3$ &           	      $1.3$ & 	       $280\times10^{2}$ &  	  $3.5\times10^{-1}$ &     	          $3$ &                    	   $120$ &           		       $121.6$ \\
				\bf{$\Comparison$}		&           	    $598.9$ &           	      $3.7$ & 	       $280\times10^{3}$ &  				   $3.5$ &     	          $3$ &                    	   $120$ &           		       $127.2$ \\
										&           	     $5995$ &           	     $30.5$ & 	       $280\times10^{4}$ &  	   $3.5\times10^{1}$ &     	          $3$ &                    	   $120$ &           		       $185.5$ \\
										&           	    $60666$ &           	    $555.7$ & 	       $280\times10^{5}$ &  	   $3.5\times10^{2}$ &     	          $3$ &                    	   $120$ &           		        $1026$ \\
										&          		   $607179$ &          			 $5349$ &      	   $280\times10^{6}$ & 		   $3.5\times10^{3}$ &       		  $3$ &               		   $120$ &               		    $8969$ \\ \hline
										&           	      $2.5$ &           	      $1.2$ & 	          		  $1752$ &  	 $2.19\times10^{-2}$ &     	          $4$ &                    	   $160$ &           		       $161.2$ \\
										&           	     $19.8$ &           	      $1.3$ & 	      $1752\times10^{1}$ &  	 $2.19\times10^{-1}$ &     	          $4$ &                    	   $160$ &           		       $161.5$ \\
										&           	    $189.7$ &           	      $2.4$ & 	      $1752\times10^{2}$ &  	    		  $2.19$ &     	          $4$ &                    	   $160$ &           		       $164.6$ \\
				\bf{$3\texttt{-}\Max$}	&           	     $1947$ &           	     $12.0$ & 	      $1752\times10^{3}$ &  	  $2.19\times10^{1}$ &     	          $4$ &                    	   $160$ &           		       $193.9$ \\
										&           	    $20121$ &           	    $216.0$ & 	      $1752\times10^{4}$ &  	  $2.19\times10^{2}$ &     	          $4$ &                    	   $160$ &           		       $595.0$ \\
										&           	   $199728$ &           	     $2415$ & 	      $1752\times10^{5}$ &  	  $2.19\times10^{3}$ &     	          $4$ &                    	   $160$ &           		        $4765$ \\
										&          		  $1976891$ &      		    	$22868$ &		  $1752\times10^{6}$ & 		  $2.19\times10^{4}$ &       		  $4$ &               		   $160$ &               		   $44928$ \\
			\end{tabular}
			\vspace{2mm}
			\caption{Evaluation of our protocols over $\Z_{2^{16}}$ with $1$ to $10^6$ batch (from the top to the bottom).}
			\label{table:appendix_protocol16}
		\end{tiny}
	\end{center}
\end{table*}
\begin{table*}[t]
	\begin{center}
		\begin{tiny}
			\begin{tabular}{c||r|r|r|r|r|r|r}
										&            \bf{pre-comp.} &         \bf{online comp.} & 		  \bf{$\#$ of comm.} &      	\bf{data trans.} & \bf{$\#$ of comm.} &                   \bf{comm.} & 				 \bf{online total} \\
										& \bf{time ($\mathrm{ms}$)} & \bf{time ($\mathrm{ms}$)} & \bf{bits ($\mathrm{bit}$)} & \bf{time ($\mathrm{ms}$)} &        \bf{rounds} & \bf{latency ($\mathrm{ms}$)} & \bf{exec. time ($\mathrm{ms}$)} \\ \hline
										&           	      $2.5$ &         	 	      $1.4$ & 	          		    $72$ &  	  $9.0\times10^{-4}$ &     	          $2$ &                    	    $80$ &           		        $81.4$ \\
										&           	     $12.7$ &         	 	      $1.5$ & 	        $72\times10^{1}$ &  	  $9.0\times10^{-3}$ &     	          $2$ &                    	    $80$ &           		        $81.5$ \\
										&           	    $112.9$ &         	 	      $1.8$ & 	        $72\times10^{2}$ &  	  $9.0\times10^{-2}$ &     	          $2$ &                    	    $80$ &           		        $81.9$ \\
				\bf{$\Equality$}		&           	     $1152$ &         	 	      $4.7$ & 	        $72\times10^{3}$ &  	  $9.0\times10^{-1}$ &     	          $2$ &                    	    $80$ &           		        $85.6$ \\
										&           	    $11404$ &         	 	     $53.4$ & 	        $72\times10^{4}$ &  				   $9.0$ &     	          $2$ &                    	    $80$ &           		       $142.4$ \\
										&           	   $114999$ &         	 	    $658.8$ & 	        $72\times10^{5}$ &  	   $9.0\times10^{1}$ &     	          $2$ &                    	    $80$ &           		       $828.8$ \\
										&          		  $1156086$ &          			 $7862$ &      	 	$72\times10^{6}$ & 		   $9.0\times10^{2}$ &       		  $2$ &                			$80$ &                		    $8842$ \\ \hline
										&           	     $22.0$ &           	      $5.8$ & 	          		  $1900$ &  	 $2.38\times10^{-2}$ &     	          $3$ &                    	   $120$ &           		       $125.8$ \\
										&           	    $183.3$ &           	      $6.3$ & 	      $1900\times10^{1}$ &  	 $2.38\times10^{-1}$ &     	          $3$ &                    	   $120$ &           		       $126.5$ \\
										&           	     $1819$ &           	     $11.6$ & 	      $1900\times10^{2}$ &  				  $2.38$ &     	          $3$ &                    	   $120$ &           		       $134.0$ \\
				\bf{$\Comparison$}		&           	    $18894$ &           	     $74.4$ & 	      $1900\times10^{3}$ &  	  $2.38\times10^{1}$ &     	          $3$ &                    	   $120$ &           		       $218.2$ \\
										&           	   $186987$ &           	    $975.0$ & 	      $1900\times10^{4}$ &  	  $2.38\times10^{2}$ &     	          $3$ &                    	   $120$ &           		        $1333$ \\
										&           	  $1861936$ &           	    $13418$ & 	      $1900\times10^{5}$ &   	  $2.38\times10^{3}$ &     	          $3$ &                    	   $120$ &           		       $15918$ \\
										&          		 $19098870$ &          		   $245178$ &      	  $1900\times10^{6}$ & 		  $2.38\times10^{4}$ &       		  $3$ &               		   $120$ &               		  $269098$ \\ \hline
										&           	     $58.5$ &           	      $6.3$ & 	      			  $9348$ &  	 $1.17\times10^{-1}$ &     	          $4$ &                    	   $160$ &           		       $166.4$ \\
										&           	    $543.7$ &           	      $8.3$ & 	      $9348\times10^{1}$ &  	    		  $1.17$ &     	          $4$ &                    	   $160$ &           		       $169.5$ \\
										&           	     $5487$ &           	     $23.7$ & 	      $9348\times10^{2}$ &  	  $1.17\times10^{1}$ &     	          $4$ &                    	   $160$ &           		       $195.4$ \\
				\bf{$3\texttt{-}\Max$}	&           	    $56608$ &           	    $270.0$ & 	      $9348\times10^{3}$ &  	  $1.17\times10^{2}$ &     	          $4$ &                    	   $160$ &           		       $547.0$ \\
										&           	   $564780$ &           	     $3440$ & 	      $9348\times10^{4}$ &  	  $1.17\times10^{3}$ &     	          $4$ &                    	   $160$ &           		        $4770$ \\
										&           	  $5721530$ &           	    $42420$ & 	      $9348\times10^{5}$ &  	  $1.17\times10^{4}$ &     	          $4$ &                    	   $160$ &           		       $54280$ \\
										&          			    $-$ &      		    	    $-$ &		        	     $-$ & 		    		     $-$ &       		  $-$ &               		     $-$ &               		       $-$ \\
			\end{tabular}
			\vspace{2mm}
			\caption{Evaluation of our protocols over $\Z_{2^{64}}$ with $1$ to $10^6$ batch (from the top to the bottom).
					 We could not execute $3\texttt{-}\Max$ with $1000000$ batch in our experiments because of the memory shortage.}
			\label{table:appendix_protocol64}
		\end{tiny}
	\end{center}
\end{table*}

\end{document}